\documentclass[aps,prd,superscriptaddress, twocolumn,nofootinbib,showpacs,floats,floatfix]{revtex4-2}
\usepackage{tensor}
\usepackage{enumitem,amssymb}
\usepackage{ragged2e}
\usepackage{graphicx}
\usepackage{comment}
\usepackage{booktabs}
\usepackage{subfig}
\usepackage{array}
\usepackage{lipsum}
\usepackage{float}
\usepackage{array}
\usepackage{verbatim}

\usepackage[dvipsnames, usenames]{xcolor}
\definecolor{linkcolor}{rgb}{0.0,0.3,0.5}
\usepackage[hypertexnames=false, unicode, colorlinks=true, linkcolor=linkcolor,
citecolor=linkcolor, filecolor=linkcolor,urlcolor=linkcolor,
pdfusetitle]{hyperref}

\usepackage[all]{hypcap}
\usepackage{graphicx}

\usepackage{bm} 
\newcommand{\fpeak}{f_{\rm peak}}
\newcommand{\fwidth}{f_{\rm width}}
\newcommand{\dApeak}{\delta \mathcal{A}_{\rm peak}}
\newcommand{\dphipeak}{\delta \phi_{\rm peak}}
\usepackage{siunitx}
\usepackage{physics}
\usepackage{orcidlink}
\usepackage{subcaption}
\usepackage[justification=raggedright,singlelinecheck=false]{caption}

\usepackage{soul}

\begin{document}


\title{Impact of detector calibration accuracy on black hole spectroscopy}

\author{Mallika R. Sinha\orcidlink{0009-0008-0906-6328}}
\email[]{mallika.sinha@monash.edu}
\affiliation{OzGrav-ANU, Centre for Gravitational Astrophysics, Research School of Physics and Research School of Astronomy \& Astrophysics, The Australian National University, Australian Capital Territory 2601, Australia}
\affiliation{OzGrav-Monash, School of Physics and Astronomy, Monash University, Clayton 3800, Australia}

\author{Ling Sun\orcidlink{0000-0001-7959-892X}}
\email[]{ling.sun@anu.edu.au}
\affiliation{OzGrav-ANU, Centre for Gravitational Astrophysics, Research School of Physics and Research School of Astronomy \& Astrophysics, The Australian National University, Australian Capital Territory 2601, Australia}

\author{Sizheng Ma\orcidlink{0000-0002-4645-453X}}
\email[]{sma2@perimeterinstitute.ca}
\affiliation{Perimeter Institute for Theoretical Physics, Waterloo, ON N2L2Y5, Canada}

\date{\today}

\begin{abstract}
Systematic errors in detector calibration can bias signal analyses and potentially lead to incorrect interpretations suggesting violations of general relativity. In this study, we investigate how calibration systematics affect black hole (BH) spectroscopy, a technique that uses the quasinormal modes (QNMs) emitted during the ringdown phase of gravitational waves (GWs) to study remnant BHs formed in compact binary coalescences. We simulate a series of physically motivated, tunable calibration errors and use them to intentionally miscalibrate numerical relativity waveforms. We then apply a QNM extraction method---the rational QNM filter---to quantify the impact of these calibration errors. We find that current calibration standards (errors within $10\%$ in magnitude and $10^\circ$ in phase across the most sensitive frequency range of 20--2000 Hz) are adequate for BH ringdown analyses with existing observations, but insufficient for the accuracy goals of future upgraded and next-generation observatories. Specifically, we show that for events with a high ringdown signal-to-noise ratio of $\sim 120$, calibration errors must remain $\lesssim 4\%$ in magnitude and $\lesssim 4^\circ$ in phase to avoid introducing biases. While this analysis focuses on a particular aspect of BH spectroscopy, the results offer quantitative benchmarks for calibration standards crucial to fully realize the potential of precision tests of general relativity in the next-generation detector era.
\end{abstract}

\maketitle
\section{\label{sec:introduction} Introduction}

Over the first three gravitational-wave (GW) observing runs by the ground-based detector network involving Advanced LIGO (aLIGO)~\cite{Aasi_2015}, Virgo~\cite{Acernese_2014}, and KAGRA~\cite{akutsu2020overview}, 90 compact binary coalescences (CBCs) have been observed \cite{O1O2, O3a, O3a.1, O3b}. The ongoing fourth observing run has yielded an additional couple of hundred significant candidates~\cite{O4-detection-rate}. The next generation of ground-based GW observatories, e.g., Cosmic Explorer~\cite{evans2021horizon} and Einstein Telescope~\cite{Maggiore2020}, will guarantee a significant number of events with high signal-to-noise ratios (SNRs), allowing for precision tests of general relativity (GR) in the strong-field regime. However, systematics from different sources, including noise artifacts, detector calibration errors, waveform systematics, and systematics arising from astrophysical origins, may lead to false indications of GR violations~\cite{Gupta2025}.  All these factors need to be thoroughly studied and quantified in order to carry out robust tests of GR.

Ground-based GW observatories operate as null experiments with a differential arm length feedback control loop. They output a time series of dimensionless strain data, used for astrophysical studies, which are reconstructed by modeling the detector response in a calibration procedure~\cite{Abbott_2017, Cahillane2017, Sun_2020, sun2021characterization, Acernese_2022,Akutsu2021}. Due to imperfect modeling and statistical variations of the detector response, as well as statistical uncertainties in the measurements, systematic errors and statistical uncertainties are present in the calibrated strain data and may affect the astrophysical analyses. Different analyses may be susceptible to different types or levels of calibration errors. Studies have been conducted to investigate the impact of calibration errors on CBC parameter estimation~\cite{Payne2020,Vitale2021,Reed2022,Wade_2023}, GW stochastic background~\cite{Yousuf2023}, measurements of the Hubble constant with bright sirens~\cite{Huang2025}, etc. However, the potential impact of calibration systematics on the variety of tests of GR that are currently employed remains to be thoroughly explored~\cite{GWTC-TGR}. Although calibration systematics may not be a limiting factor with current detectors due to the relatively low ringdown SNRs of observed events, a clear understanding of the calibration accuracy requirements is essential to establish guidelines for the calibration technologies and standards necessary for next-generation GW detectors~\cite{evans2021horizon}.
In this study, we quantify the impact of calibration systematics and derive a benchmark for calibration standards in future GW observatories by focusing on a specific aspect of testing GR---ringdown analysis via BH spectroscopy.

GWs emitted during the ringdown stage of a binary black hole (BH) merger are produced by the disturbed remnant BH ringing down into a stable state. These ringdown GW signals are predicted by the BH perturbation theory and are described by a superposition of complex-valued quasinormal modes (QNMs) \cite{Kokkotas_1999, rev_QNM_Berti_2009,Berti:2025hly,vishveshwara1970scattering, PhysRevD.34.384,chandrasekhar1975quasi, Hans-Peter_Nollert_1999}. The frequencies and damping times of these QNMs are solely determined by the intrinsic properties of the BH, mass and spin, as dictated by the no-hair theorem \cite{Carter1971}. BH spectroscopy, through the analysis of multiple individual ringdown modes, provides a means of testing the no-hair theorem~\cite{Dreyer_2004qnm2,Berti_2006qnm3,Cardoso_2016}. 
A couple of relatively high-SNR events to date have already enabled BH spectroscopy, e.g., with GW150914~\cite{Carullo_2019, Finch_2022, wang2024gatingandinpaintingperspectivegw150914ringdown, Abbott_2021_tests, Bustillo_2021, Ghosh_2021, Mode_Cle_Ma_2023, Ma_2023b, Crisostomi_2023, Wang_2025, correia2024lowevidenceringdownovertone, nohhair_Isi_2019, isi2022revisiting, Lu2025}, GW190521~\cite{Capano_2023, Abbott_2020, siegel2023ringdowngw190521hintsmultiple, capano2024estimatingfalsealarmrates}, and more~\cite{Ghosh_2021, Abbott_2021_tests, GWTC-TGR}. Note that these studies, performed with various analysis methods, do not agree on key observations such as the detection of an overtone~\cite{Cotesta_2022, Isi_2023, Carullo_2019, 2023PhRvL.131p9002C}, or a higher-order mode~\cite{Capano_2023, Abbott_2020, siegel2023ringdowngw190521hintsmultiple, capano2024estimatingfalsealarmrates}. Such ambiguity in conclusions is primarily due to the relatively low ringdown SNRs in existing events, but with future high-SNR events, it can also arise from unaccounted-for systematics that affect each individual analysis method, such as detector calibration errors. 

Here, we adopt a BH ringdown QNM analysis method, the QNM rational filter \cite{Mode_Cle_Ma_2023, QNMF_Ma_2022, Ma_2023b}. We simulate BH ringdown signals using numerical relativity (NR) waveforms from the Simulating eXtreme Spacetimes (SXS) catalogue \cite{sxs,Scheel:2025jct, sxs_nodate} and evaluate the recovery of the BH properties via QNM analysis with and without the existence of calibration errors in the data. To simulate different scenarios of calibration errors, a set of physically motivated, parameterized, synthetic calibration errors is generated, with reference to the real calibration errors witnessed in the aLIGO detectors during the third observing run. In addition, detector noise present in real observations affects the analysis even if the data are perfectly calibrated. We step-by-step evaluate the impact of calibration errors by analyzing pure BH ringdown waveforms with zero noise (i.e., miscalibrating the waveform itself), simulated signals in white Gaussian noise, and simulated signals in coloured Gaussian noise at the design sensitivity of existing and future detectors. The results quantify the calibration requirements for GW detectors necessary to ensure robust analyses.

The structure of the paper is as follows.
In Sec.~\ref{sec:error}, we review the GW detector calibration procedure and typical calibration error profiles, and introduce a method to simulate realistic calibration errors.
In Sec.~\ref{sec:method}, we review the rational filter as a BH spectroscopy method and investigate the impact of calibration errors on QNM analysis using NR simulations. The calibration accuracy standards required in existing and future-generation observatories are presented in Sec.~\ref{sec:requirements}. Finally, we conclude and discuss future work in Sec.~\ref{sec:conclusion}.

\section{\label{sec:error} Calibration Systematics}
\subsection{Detector calibration}
Ground-based GW detectors output a digital voltage error signal, $d_{\rm err}$, under the differential arm length feedback control loop. Dimensionless strain time-series data---used for all astrophysical analyses---are reconstructed from the raw digital error signal using a complex-valued frequency-dependent transfer function, built from the modeled detector response. The procedure of reconstructing the strain data is called detector calibration, and the modeled detector response is referred to as a calibration model \cite{Abbott_2017, Cahillane2017, Sun_2020,sun2021characterization,Acernese_2022,Akutsu2021}. 

The strain data, $h$, is defined by the differential length changes in the two arms, $\Delta L_{\rm free}$, divided by the arm length, $L$, i.e., $h\equiv{\Delta L_{\rm free}}/{L}$.
Note here that the strain data $h$ is not the GW strain, but the measured strain in the detector, which contains both noise and, potentially, GW signals. However, $\Delta L_{\rm free}$ is not directly measurable given the differential arm length is controlled. The feedback loop suppresses changes in the differential arm length and leaves only a residual differential arm displacement, $\Delta L_{\text{res}}$. The measured analogue residual displacement, $\Delta L_{\rm res}$, is converted to the digital output error signal, $d_{\rm err}$, through a sensing transfer function, $C$:\footnote{Here, we describe the derivation fully in the frequency domain.}
\begin{equation}
  d_{\rm err}= C\Delta L_{\text{res}}.
\end{equation}
A set of digital filters, $D$, are then applied to $d_{\rm err}$ to produce the digital control signal:
\begin{equation}
    d_{\rm ctrl} = D d_{\rm err}, 
\end{equation}
which is fed back through the actuation path. This produced an analog control displacement, $\Delta L_{\rm ctrl}$, on the test masses to suppress $\Delta L_{\text{free}}$:
\begin{equation}
    \Delta L_{\text{ctrl}}= A d_{\rm ctrl},
\end{equation}
where $A$ denotes the actuation transfer function. With $\Delta L_{\text{ctrl}}$ subtracted from $\Delta L_{\text{free}}$, all that remains is the residual $\Delta L_{\text{res}}$. Combining the above functions, we find:
\begin{equation} 
    \begin{split}
        \Delta L_{\text{free}}
        & = \left[ \frac{1}{C} + AD\right] d_{\rm err} = R\,d_{\rm err}.
    \end{split}
\end{equation}
Here, $R \equiv 1/C +AD$ is the detector response function \cite{Abbott_2017,Sun_2020}, such that we can write the strain in terms of the readout error signal:
\begin{equation}
  h= \frac{R\,d_{\rm err}}{L}.
  \label{eq:h_vs_R}
\end{equation}
In the calibration procedure, since the true frequency-dependent, time-dependent response function $R$ is not perfectly known, $h$ is reconstructed by a modeled response function, $R^{\rm (model)}$. 
Calibration systematic errors arise from imperfections in the modeled response function, $R^{\rm (model)}$, due to imperfect estimates of model parameters, uncompensated time dependence, or additional features in the true response function that are missing in the model. Given $d_{\rm err}$ is a digital signal and $L$ is measured to high precision [Eq.~\eqref{eq:h_vs_R}], the errors in the response are equivalent to the errors in the reconstructed stain data.

We can write the frequency-dependent and time-dependent calibration errors as a ratio, $\eta(f,t)$, between the true response and modeled response functions:
\begin{equation}
    \eta(f,t)= [1+\delta \mathcal{A}(f,t) ]e^{i \delta \phi(f,t)}=\frac{R(f,t)}{R^{\rm (model)}(f,t)},
\end{equation}
where $\delta \mathcal{A}(f,t)$ and $\delta \phi(f,t)$ denote the magnitude and phase errors, respectively. A perfect model corresponds to $\eta =1$, i.e., $\delta \mathcal{A} = 0$ and $\delta \phi =0$. In this study, we focus on the impacts of calibration errors on individual CBC signals, which last less than a second. As such, we assume that the slow-varying time dependence (on the order of hours to days~\cite{Sun_2020,sun2021characterization}) has negligible effects and approximate these calibration errors as frequency-dependent, but constant in time, viz.
\begin{equation} 
\label{eq:Cerreq}
    \eta(f) =\qty[1+\delta \mathcal{A}(f)]e^{i\delta \phi(f)} = \frac{R(f)}{R^{\rm (model)}(f)}.
\end{equation}

Regular detector measurements are used to estimate potential calibration errors present in the strain data and can be accounted for when performing certain parameter estimation techniques \cite{Sun_2020, O3b, O3a, O1O2, Aasi2013,Abbott2016GW150914PE}. 
However, some calibration systematics may exist without being identified in error estimation. Additionally, in certain analyses, the estimates of calibration errors are not fully taken into consideration, since they are not the dominant source of uncertainty at this stage. 

It is generally accepted that calibration errors do not tend to introduce biases in GW analyses with current-generation detectors \cite{Wade_2023, Huang2025, Payne2020, Vitale2021,Yousuf2023}, and the uncertainty in astrophysical parameter estimation is instead dominated by noise fluctuations due to the limited SNR that can be achieved with current detector sensitivities. The situation is likely to change when significantly increased SNRs can be achieved in future upgraded detectors and next-generation observatories. The impact of calibration errors on future GW data analyses has yet to be comprehensively evaluated, with \citet{evans2021horizon} predicting that a limit of 1\% in magnitude, i.e., $\delta \mathcal{A} < 1\%$ is needed for next-generation observatories, such as Cosmic Explorer, in order to allow robust scientific analyses. For reference, during the third observing run (O3), the upper limits on the calibration systematic errors (with associated uncertainties) for the two aLIGO detectors are $\lesssim10$\% in magnitude and $\lesssim10^\circ$ in phase, at a 1-$\sigma$ confidence interval in the most sensitive frequency band of 20--2000~Hz~\cite{Sun_2020,sun2021characterization}.

\subsection{Calibration error profiles}
\label{sec:error_profile}
\begin{figure*}
    \includegraphics[width=0.6\linewidth]{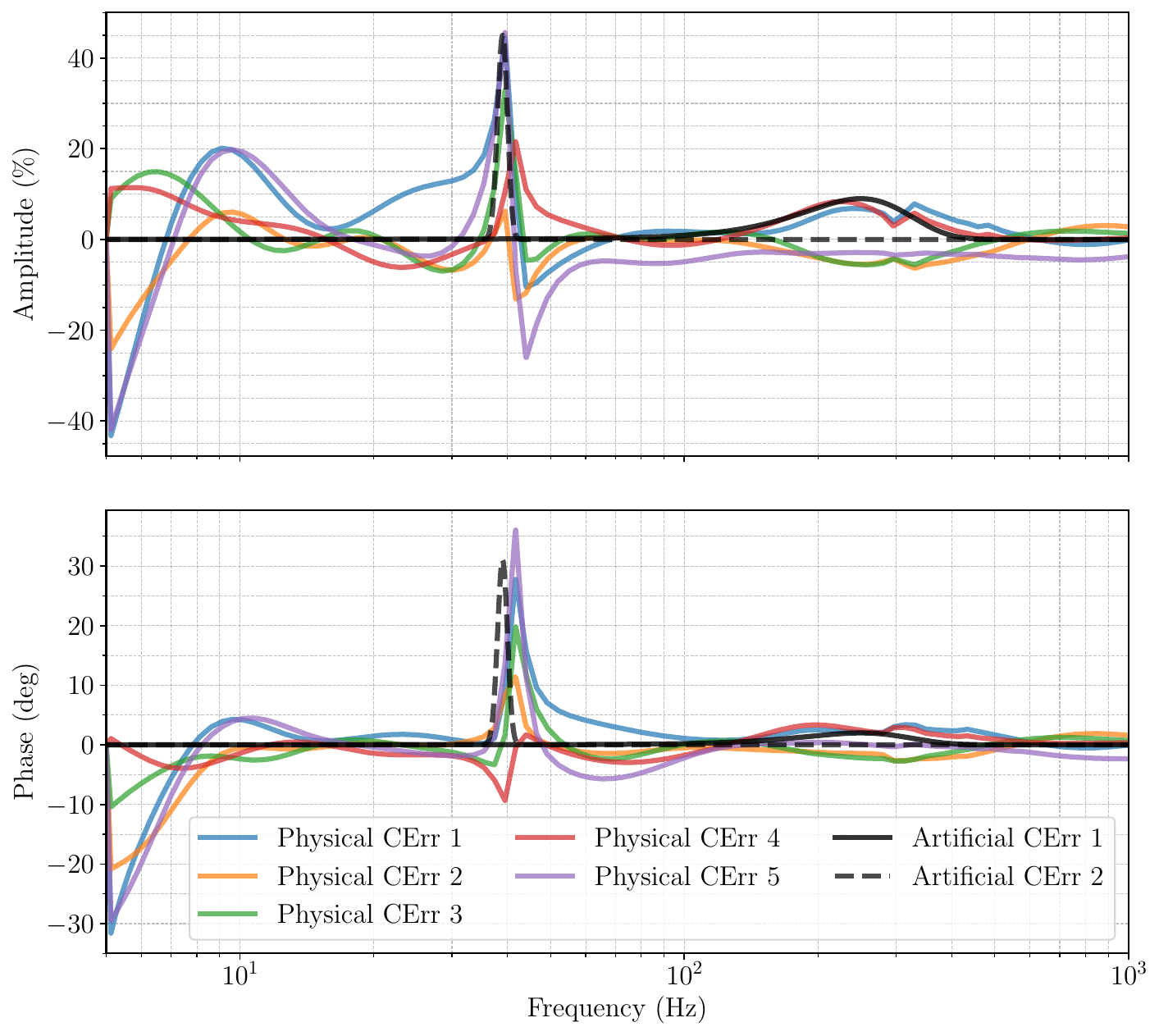}
    \caption{\label{fig:some_real} Examples of realistic physical calibration errors (labelled as `Physical CErr') estimated during O3 at aLIGO Livingston for a given outlier time (colored)~\cite{L1cerr,sun2021characterization} and examples of artificial calibration errors (labelled as `Artificial CErr') generated using a parameterized approach in this study (black). The solid and dashed black curves correspond to a broad error and a sharp error, respectively; see text for detailed parameters.}
\end{figure*}

Calibration errors can affect the amplitude and/or phase of the observed waveform [see Eq.~\eqref{eq:Cerreq}] as a multiplicative effect on the `true' strain in the frequency domain. The characterization of calibration systematics has been conducted in all existing observing runs with aLIGO \cite{Abbott_2017, Cahillane2017, Sun_2020,sun2021characterization, cerrcurves} and Virgo \cite{Acernese_2022, cerrcurves}. In this study, we refer to the calibration errors seen in the latest aLIGO O3 data \cite{Sun_2020,sun2021characterization}. In Fig.~\ref{fig:some_real}, we plot five out of $10^4$ numerically estimated samples of the calibration systematic error at an outlier time (relatively larger calibration error and uncertainty) at the aLIGO Livingston detector, shown as colored curves~\cite{Sun_2020,sun2021characterization}. We can see some characteristic features in these examples: errors may peak at some frequencies, spanning a wide band, or appearing as a sharp spike in a narrow frequency band, depending on the physical causes of the error, e.g., missing features in pendulum dynamics, an imperfect compensation of the electronics, or unaccounted for thermalization effects~\cite{Sun_2020,sun2021characterization}. 

We parameterize the calibration errors using four main factors: 1) the frequency at which the frequency-domain error is at a maximum, $f_{\text{peak}}$, 2) the characteristic width of the frequency-domain error, $f_{\text{width}}$, 3) the peak of the magnitude error, $\delta \mathcal{A}_{\rm peak}$, and 4) the peak of the phase error, $\delta \phi_{\rm peak}$. With these four factors, we can simulate tunable artificial calibration errors that mimic the realistic errors with physical origins. Most realistic physical errors, as shown in Refs.~\cite{Sun_2020,sun2021characterization} and the examples in Fig.~\ref{fig:some_real}, have multiple features across the full frequency band with wildly varied shapes. Here, we only simulate one feature at a time since the analysis of QNMs usually does not span the full frequency range, but rather is focused on a band of $\sim 10$--$500$~Hz; calibration errors far from the QNM frequencies do not have an impact on the analysis (see Sec.~\ref{sec:nonoise}).\footnote{We leave in-depth analyses of more complex calibration systematics, including studies with real calibration errors, to future work.} The shape of an individual feature can be approximated by a Gaussian function, which has smooth edges and thus naturally prevents the introduction of artifacts to the time-domain data.\footnote{The simulated errors do not mimic some of the realistic features such as those crossing unity magnitude and zero phase, e.g., the features around 40~Hz in the examples shown in Fig.~\ref{fig:some_real}. Such features can be regarded as a combination of two errors in opposite directions. Given that this study aims to quantify the impact of calibration errors in general situations, we do not consider specific combinations of features here.}

We define the width of the Gaussian-shaped artificial error to span $\pm 2 \sigma$ around the center frequency. The frequency-domain error function in Eq.~\eqref{eq:Cerreq} is constructed with:
\begin{equation}
\begin{split}
    \delta \mathcal{A}(f) &= \delta \mathcal{A}_{\rm peak} \exp\left[ -\frac{1}{2}\qty(\frac{f-f_{\text{peak}}}{f_\text{width}/4})^2 \right], \\
    \delta \phi(f) &= \delta \phi_{\rm peak} \exp\left[ -\frac{1}{2}\qty(\frac{f-f_{\text{peak}}}{f_\text{width}/4})^2\right].
\end{split}
\end{equation}
We plot two examples of the simulated artificial errors in Fig.~\ref{fig:some_real}, one with $f_{\rm peak} = 39$~Hz, $f_{\rm width} = 4$~Hz, $\delta \mathcal{A}_{\rm peak} = 45\%$, and $\delta \phi_{\rm peak}=31^\circ$ (dashed black curve), and the other with $f_{\rm peak} = 250$~Hz, $f_{\rm width} = 280$~Hz, $\delta \mathcal{A}_{\rm peak} = 9\%$, and $\delta \phi_{\rm peak}=8^\circ$ (solid black curve). It is physically motivated to set the magnitude and phase error peaks at the same frequency because any imperfect modeling in the calibration procedure is likely to produce both a magnitude and phase error. Fig.~\ref{fig:some_real} demonstrates that these simulated errors can sufficiently capture the features of realistic physical calibration errors (shown in the frequency domain).

\section{\label{sec:method} Impact of calibration errors on quasinormal mode analysis}

We start by reviewing QNMs and the rational filter analysis method (Sec.~\ref{sec:qnm}). We then investigate how calibration errors impact pure signal waveforms (Sec.~\ref{sec:nonoise}) and discuss the situation with additive noise (Sec.~\ref{sec:white_noise}). 

\subsection{\label{sec:qnm} Quasinormal modes and the rational filter}

At the linear order, the GWs emitted during the ringdown phase can be described by the superposition of a series of complex-valued QNMs~\cite{Kokkotas_1999, rev_QNM_Berti_2009,Berti:2025hly,vishveshwara1970scattering, PhysRevD.34.384,chandrasekhar1975quasi, Hans-Peter_Nollert_1999}. 
The QNMs are labeled by two angular index numbers, $(\ell,m)$, and an overtone index, $n$. For example, the fundamental (zeroth overtone) $\ell=2,\ m = 2$ mode is referred to as the 220 QNM, which is considered the dominant mode for quasi-circular binary systems~\cite{Dreyer_2004qnm2, Flanagan_1998}. While the emission strength of the QNMs are contingent on the excitation source\footnote{In this case, we would consider the initial BBH system to be the excitation source.} and the inherent angular distribution, the complex-valued frequency component for each $\ell m n$ mode, $\omega_{\ell m n}$, is exclusively determined by the intrinsic parameters of the remnant BH, mass and spin~\cite{Carter1971, Kokkotas_1999, rev_QNM_Berti_2009,Berti:2025hly}. BH spectroscopy is used to determine the remnant BH properties using solely $\omega_{\ell mn}$, offering a consistency test for GR when compared to the results obtained from the full inspiral-merger-ringdown analyses~\cite{GWTC-TGR}. Calibration systematics can distort QNM waveforms, potentially leading to false indications of GR violations and introducing biases in BH parameter estimation. True deviations from GR are most likely to appear as subtle discrepancies between analyses---a pattern that can also be caused by calibration errors when only one analysis method accounts for them. 

In this study, we leverage a QNM analysis method, the rational QNM filter~\cite{QNMF_Ma_2022,Mode_Cle_Ma_2023,Ma_2023b, Lu2025}, to study the impact of calibration errors on BH ringdown analyses. The filter, designed to reveal sub-dominant modes by cleaning (i.e., subtracting) dominant modes from the ringdown, provides a robust tool for BH spectroscopy. We can remove a set of linearly superpositioned QNMs using a total rational QNM filter applied in the frequency domain
\begin{equation}
    \mathcal{F}_{\rm tot}=\prod_{\ell mn} \mathcal{F}_{\ell mn}=\prod_{\ell mn}\frac{\omega-\omega_{\ell mn}}{\omega-\omega_{\ell mn}^*}\frac{\omega+\omega_{\ell mn}^*}{\omega+\omega_{\ell mn}},
\end{equation}
where $\mathcal{F}_{\ell mn}$ is the filter for each $\ell mn$ QNM, and $^*$ denotes the complex conjugate. Note that $\mathcal{F}_{\ell mn}$ is entirely determined by the QNM frequency $\omega_{\ell mn}$, which is, in turn, determined by the remnant BH mass and spin. When applying a filter $\mathcal{F}_{\ell mn}$ corresponding to the $(\ell, m, n)$ mode, the amplitudes of other QNMs with $(\ell',m',n')\neq (\ell, m, n)$ are reduced by a factor of $B_{\ell mn}^{\ell'm'n'}$ depending on the $(\ell',m',n')$ mode frequency and the phase between the two modes. (See Ref.~\cite{Ma_2023b} for the full derivation of $B_{\ell mn}^{\ell'm'n'}$.) 

Suppose all the QNMs are removed with a total filter built from the true BH parameters, the residual is consistent with pure noise.
By evaluating the residual in a Bayesian framework, we can estimate the mass and spin of the final BH, denoted by $M$ and $\chi$.\footnote{Throughout this study, we evaluate the BH mass and spin in the detector frame.}
We use a likelihood function~\cite{Ma_2023b}:
\begin{equation}
    \ln P(d_t|M, \chi, t_0, \mathcal{F}_{\text{tot}}) = -\frac{1}{2}\sum_{i,j>I_0}d_i^FC_{ij}^{-1}d_j^F,
    \label{eq:likelihood}
\end{equation}
where $d_t$ denotes the time-series data being analyzed, $d_i^F$ are samples of the filtered data after $t_0$ at discrete times indexed by $i$, $I_0$ is the index for time $t_0$, and $C_{ij}$ is the noise autocovariance matrix. The likelihood function can be converted into a joint posterior of $M$ and $\chi$~\cite{Ma_2023b}:
\begin{equation} 
\begin{split}
    \ln P(M, \chi|d_t, t_0, \mathcal{F}_{\text{tot}}) &= \ln P(d_t|M, \chi, t_0, \mathcal{F}_{\text{tot}})\\& + \ln \Pi (M, \chi) + \text{constant},
    \label{eq:posterior}
\end{split}
\end{equation}
where $\Pi (M, \chi) $ is the prior. The two-dimensional joint posterior can be directly and efficiently computed through the likelihood function on a grid of $(M, \chi)$ without using a full Markov chain Monte Carlo analysis~\cite{Ma_2023b}. The rational QNM filter and this Bayesian framework have been applied to the first detection event, GW150914, identifying the existence and quantifying the significance of the first overtone 221 mode in the ringdown signal~\cite{Mode_Cle_Ma_2023,Ma_2023b, Lu2025}. 

Here, we define two quantities as the metric to evaluate the recovery of BH parameters ($M$ and $\chi$) in ringdown signals using the rational filter. First, a joint posterior quantile that can be computed as~\cite{Lu2025}:
\begin{align}
    p(M_t,\chi_t)=\frac{\sum_{\mathcal{L}(d|M,\chi)>\mathcal{L}(d | M_t,\chi_t)}\mathcal{L}(d|M,\chi)}{\sum_{M, \chi}\mathcal{L}(d|M,\chi)} \, , \label{eq:posterior_quantile}
\end{align}
where $M_t$ and $\chi_t$ are the true values of BH mass and spin, respectively. For example, $p(M_t, \chi_t)=0.9$ indicates that $(M_t, \chi_t)$ lies on the 90\% credible region contour, i.e. the interval in which $(M_t, \chi_t)$ will fall 90\% of the time. A lower $p(M_t, \chi_t)$ value indicates a more accurate recovery of $(M_t, \chi_t)$. Second, we define a parameter distance, $\epsilon$, between the maximum a posteriori (MAP) values in the joint posterior, denoted by $(M_{\rm MAP}, \chi_{\rm MAP})$, and the true BH parameters, given by~\cite{Ma_2023b}:
\begin{equation}
    \label{eq:epsilon}
    \epsilon = \sqrt{\left(\frac{M_{\rm MAP}-M_t}{M_t}\right)^2+(\chi_{\rm MAP}-\chi_t)^2}.
\end{equation}
Smaller $\epsilon$ values indicate better agreement between the best estimates of the BH parameters, $(M_{\rm MAP}, \chi_{\rm MAP})$, and the true values, $(M_t, \chi_t)$. 

\subsection{\label{sec:nonoise}Miscalibrate pure waveforms}

We start by investigating the impact of calibration errors on pure signals. We intentionally `miscalibrate' the signal, i.e., apply a calibration error to the signal waveform, and compare the results from the miscalibrated signals and those from the accurate ones. We take BBH merger numerical relativity (NR) waveforms from the Simulating eXtreme Spacetimes (SXS) catalog \cite{sxs,Scheel:2025jct, sxs_nodate} (referred to as NR waveforms onward) and `inject' the NR waveforms into the detector, i.e., convert the signal to the detector frame. The strain signal can be decomposed into the spin-weighted spherical harmonic basis, given by:
\begin{equation}
   \begin{split}
        \label{eq:qnm_NR_approx}
        h(\iota,\beta,t)&=\qty(h_+-ih_{\cross})(\iota,\beta,t) \\
        &=\sum_{\ell m}{}_{-2}Y_{\ell m}(\iota,\beta)h_{\ell m}(t),
   \end{split}
\end{equation}
where $\iota$ is the inclination angle, $\beta$ is the azimuth angle, and ${}_{-2}Y_{\ell m}$ are the spin-weighted spherical harmonics.

\begin{figure*}
    {%
        \includegraphics[width=.49\linewidth]{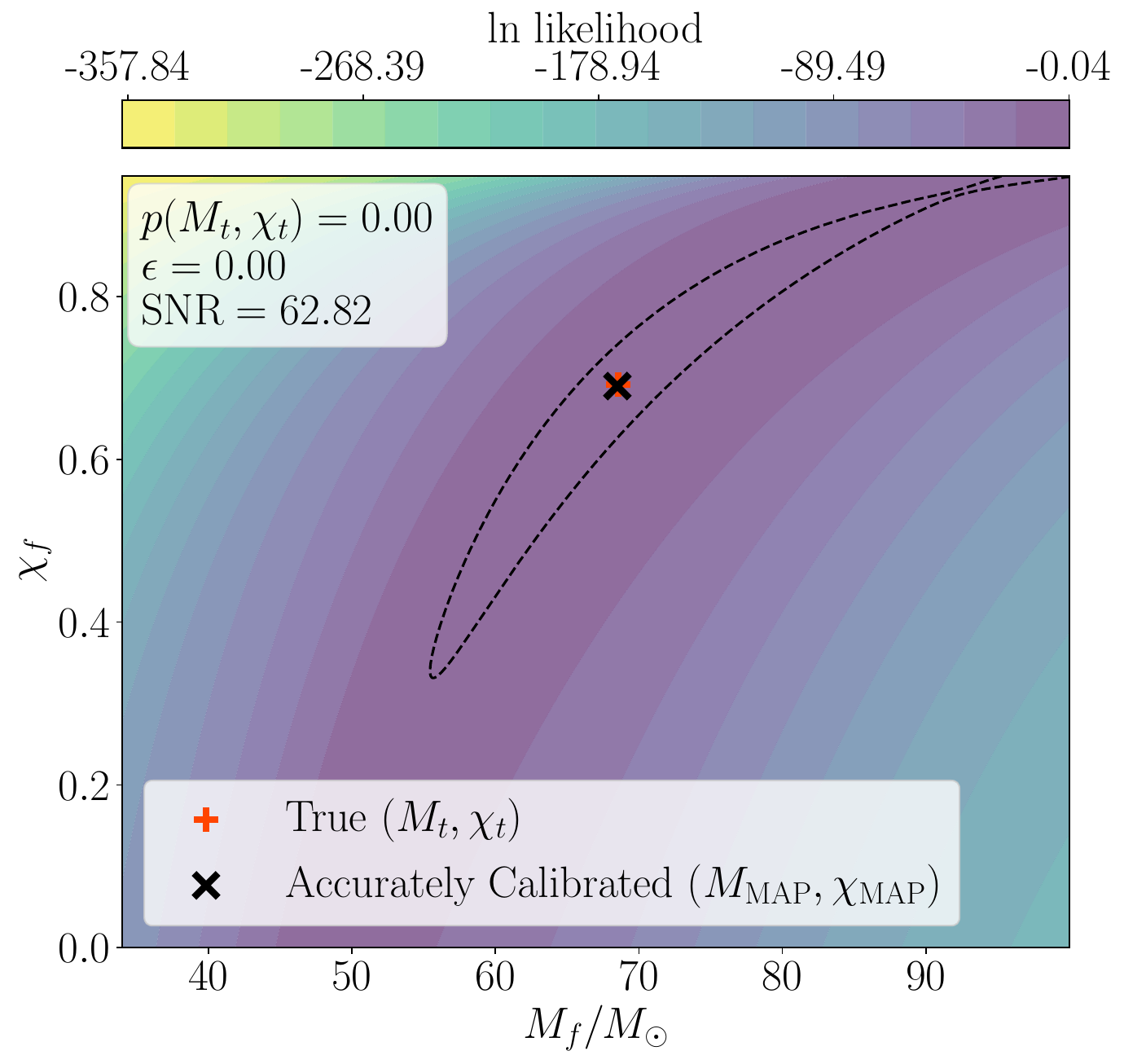}%
        \label{subfig:acc_cal}%
    }\hfill
    {%
        \includegraphics[width=.49\linewidth]{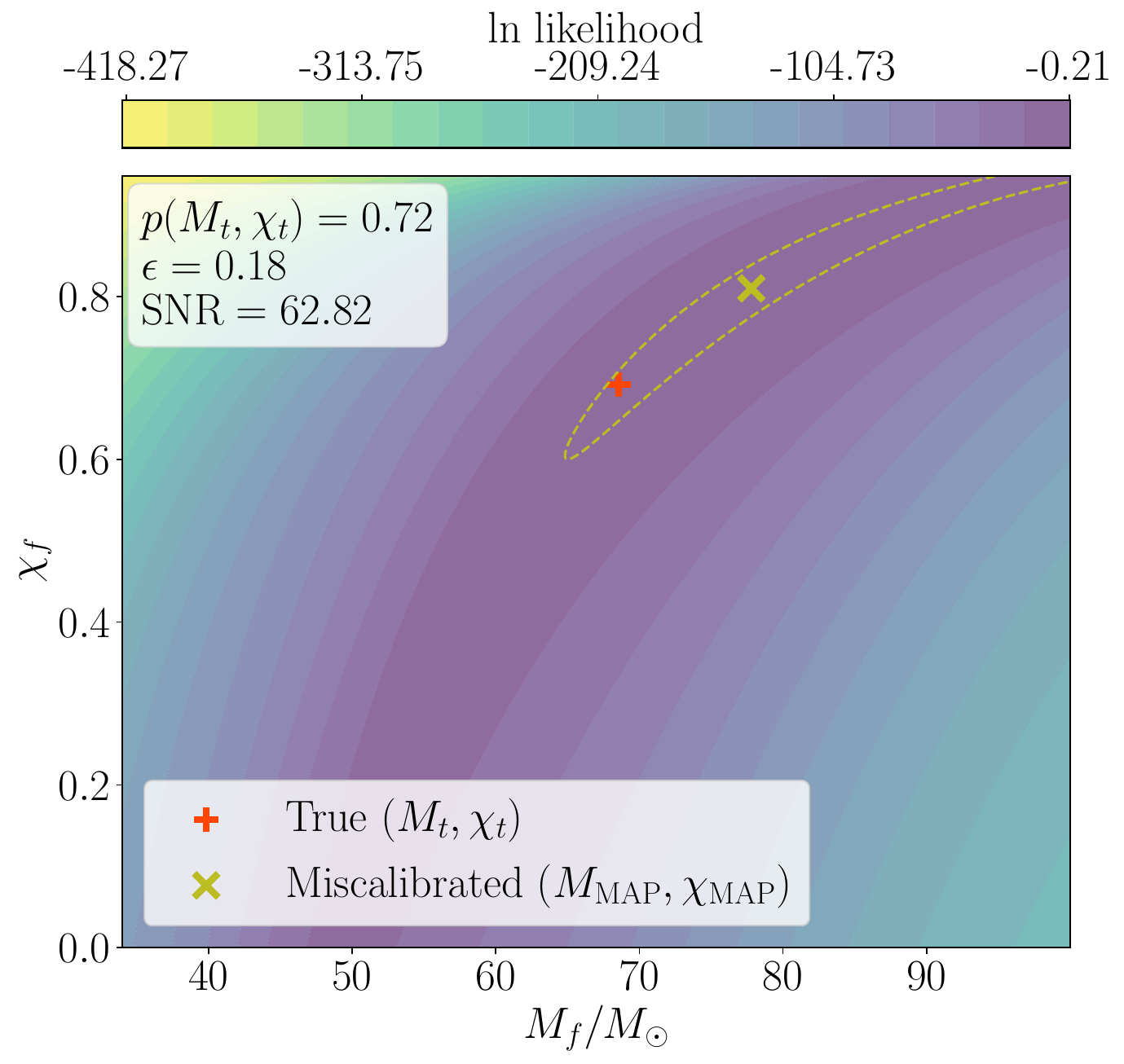}%
        \label{subfig:mis_cal}%
    }
    \caption{\label{fig:no_noise_contour}Joint posterior distribution of $(M, \chi)$ for a GW150914-like NR waveform injected into aLIGO Hanford detector frame without additive noise by applying a $\{220,221\}$ rational filter. Left: No calibration error is added. Right: The signal is miscalibrated with an error of $\dphipeak=10^\circ$ peaked at the 220 mode frequency. The dashed contour indicates the 90\% credible region obtained from the QNM rational filter analysis. The plus and cross markers indicate the true BH parameters and MAP estimates, respectively.}
\end{figure*}

\begin{figure*}[!tbh]
        \includegraphics[width=.9\linewidth]{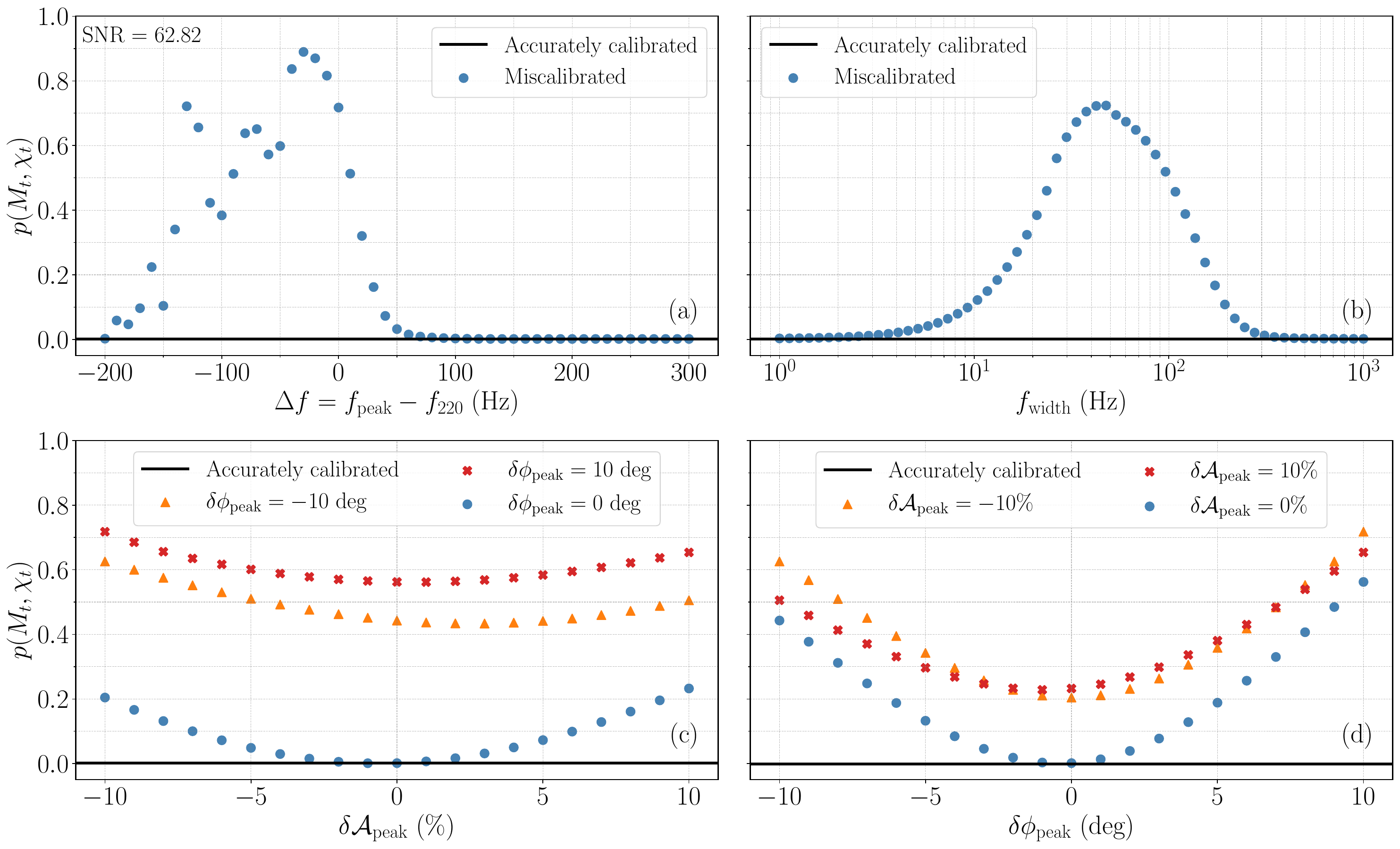}
    \caption{\label{fig:pval0305}Joint posterior quantile values, $p(M_t, \chi_t)$, for a GW150914-like NR waveform (${}_{-2}Y_{2\pm2}$ components only) injected in aLIGO Hanford without additive noise as a function of the calibration error properties: (a) frequency offset between the peak frequency of the error and the 220 QNM frequency, $\Delta f=f_{\text{peak}}-f_{220}$, where $f_{220}=249.43$ Hz ($\fwidth=50$ Hz, $\dApeak=-10\%$, and $\dphipeak=10^\circ$), (b) error width $f_{\text{width}}$ ($\Delta f=0$, $\dApeak=-10\%$, and $\dphipeak=10^\circ$), (c) peak magnitude error $\dApeak$ ($\Delta f=0$, $\fwidth=50$ Hz), and (d) peak phase error $\dphipeak$ ($\Delta f=0$, $\fwidth=50$ Hz). In (c) and (d), the colored markers indicate different fixed values of $\dphipeak$ and $\dApeak$, respectively. The black horizontal lines indicate the perfect recovery of $(M,\chi)$ from correctly calibrated data, with all $p(M_t, \chi_t)$ aligned with 0.}
\end{figure*} 

\begin{table}[!tbh]
\centering
\caption{\label{table:NR_waveforms}Remnant BH parameters for the two representative NR waveforms used in this study
~\cite{sxs,Scheel:2025jct, sxs_nodate}. The remnant mass is defined by a ratio to the total BBH mass given by the NR waveform, here we set the BBH mass to be $70.6\text{M}_\odot$~\cite{Abbott2016GW150914PE}.}
\begin{tabular}{p{0.15\textwidth} p{0.12\textwidth} p{0.1\textwidth} p{0.05\textwidth}}
\toprule
\textbf{SXS ID} & \textbf{Remnant} & \textbf{Remnant} & \textbf{Mass} \\  
 & \textbf{Mass ($M_\odot$)} & \textbf{Spin} & \textbf{Ratio} \\ 
\midrule 
SXS:BBH:0305    & 67.21 & 0.69 & 1.22 \\
SXS:BBH:0181    & 69.57 & 0.37 & 6.00 \\ 
\bottomrule       
\end{tabular}
\end{table}

In this study, we focus on two representative NR waveforms: 1) a GW150914-like system (SXS:BBH:0305) in which the $\ell=m=2$ QNMs are dominant, and 2) a higher mass-ratio ($q=6.00$) system (SXS:BBH:0181) in which the higher-order angular $\ell=m=3$ mode is observable (see key parameters in Table~\ref{table:NR_waveforms}). We filter the $\{220,221\}$ and $\{220,221,330\}$ QNMs for the analysis, respectively.
Note that the dominant mode components in the ringdown wave depend on the analysis starting time, $t_i$. Here, $t_i=16 M_t$ is chosen for the rest of this study as all the QNMs considered have sufficient amplitudes and contribute to the parameter estimation (see Appendix~\ref{app:time} for discussions about the start time). For the GW150914-like waveform, the strongest emission is towards the directions with $\iota=0$ or $\iota=\pi$. We set $\iota=0$ for injections with waveform SXS:BBH:0305. However, the $\ell=m=3$ mode does not emit towards $\iota=0$ or $\iota=\pi$. Thus, we set $\iota=0.25\pi$ for injections with waveform SXS:BBH:0181. Without loss of generality, we set all other parameters, including the source sky location, right ascension and declination, polarisation angle, and azimuth angle, to 0. We have validated that changing these parameters degenerates with tuning the SNR, which we can simply adjust by varying the luminosity distance or the noise level in a controlled way. In this section, since we are analyzing pure signal waveforms, we scale the autocovariance function (i.e., the spectral density of the signal) to produce signals with a reasonable fiducial ringdown $\text{SNR} \sim60$.\footnote{To understand any potential impact from numerical noise and mode mixing effects in NR waveforms, we first validate the method with a manually constructed well-controlled ringdown signal with only two QNMs, 220 and 221, or 220 and 330, before analyzing the NR waveforms. See Appendix~\ref{app:manual} for details.}

\begin{figure*}
    \includegraphics[width=.9\linewidth]{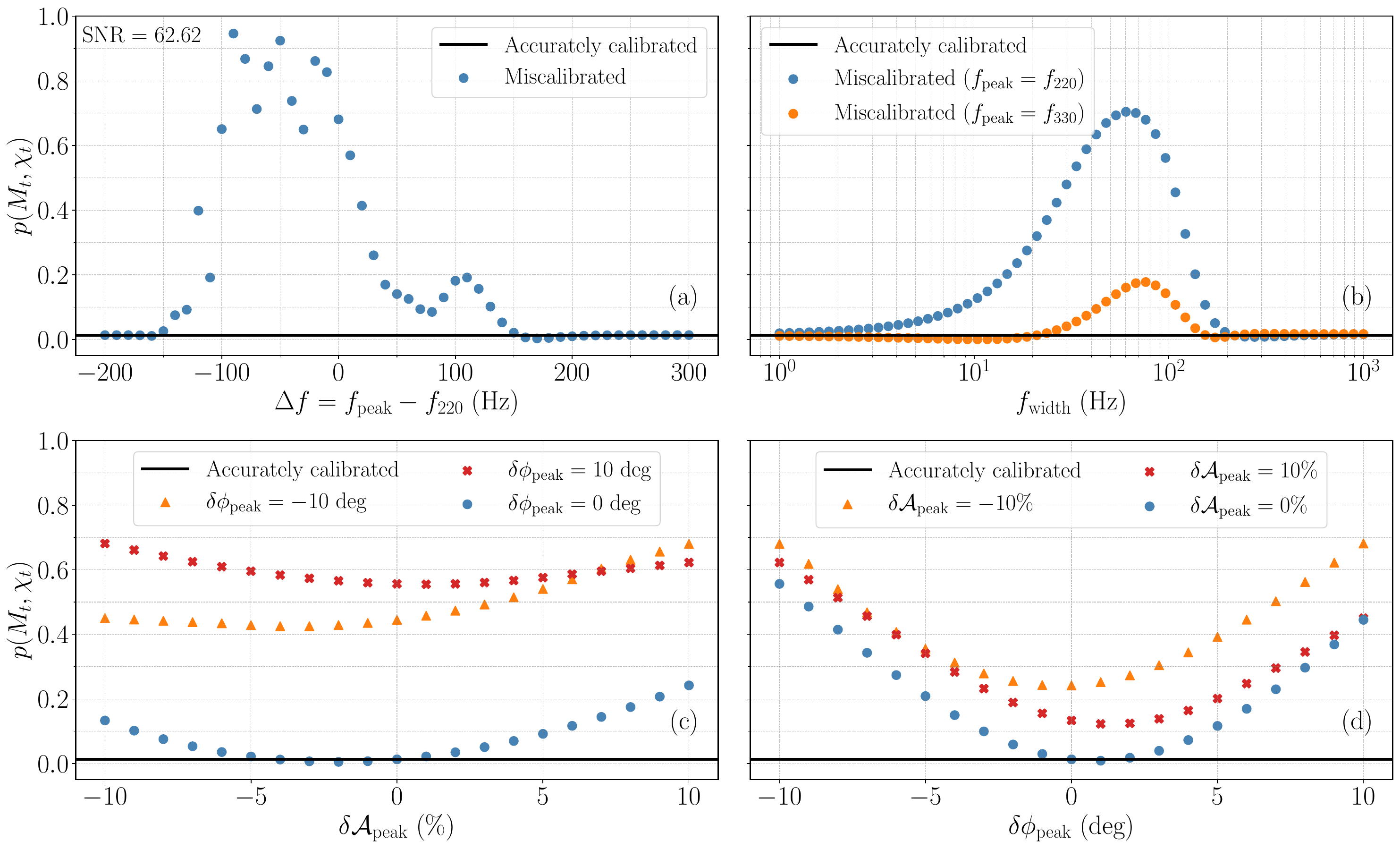}%
    \caption{\label{fig:pval330}(Similar to Fig.~\ref{fig:pval0305}) Joint posterior quantile values, $p(M_t, \chi_t)$, for a high mass ratio NR waveform (${}_{-2}Y_{2\pm2}$ and ${}_{-2}Y_{3\pm3}$ components only) injected in aLIGO Hanford without additive noise as a function of the calibration error properties. Here, the QNM filter includes the 220, 221, and 330 QNMs. In panel (b), both scenarios of $f_{\rm peak}$ = $f_{220}$ and $f_{\rm peak}$ = $f_{330}$ are tested (with fixed $\dApeak=-10\%$ and $\dphipeak=10^\circ$).
    }
\end{figure*}

In Fig.~\ref{fig:no_noise_contour}, we show an example of the impact of miscalibration. The joint posterior distributions of the BH mass and spin are obtained for a GW150914-like NR waveform (SXS:BBH:0305, ${}_{-2}Y_{2\pm2}$ components only) using a rational filter that consists of the $\{220,221\}$ modes. The signal is injected into aLIGO Hanford detector frame without additive noise. We compare the posteriors obtained from the rational filter without (left) and with (right) a calibration error. The error is set to be centered at the 220 QNM frequency ($f_{220}=249.43$~Hz) with a peak phase error of $\delta \phi_{\rm peak} = 10^\circ$ ($\fwidth=50$~Hz, $\dApeak=0\%$). Note that the lower the $p(M_t, \chi_t)$ value, the better the recovery of the BH parameters. The orange plus represents the true values, $(M_t, \chi_t)$, the dashed curve marks the 90\% credible region contour, and the cross indicates the MAP estimates, $(M_{\rm MAP}, \chi_{\rm MAP})$. In the accurately calibrated case, we have $(M_t, \chi_t)=(M_{\rm MAP}, \chi_{\rm MAP})$ and thus $p(M_t, \chi_t)=\epsilon = 0$, indicating a perfect recovery when the data is accurately calibrated and there is no additive noise. This is not the case on the right with the results from miscalibrated data shifted away from $(M_t, \chi_t)$, yielding a non-zero $p(M_t, \chi_t)$ and $\epsilon$. 

In Fig.~\ref{fig:pval0305}, we show the joint posterior quantile values for various calibration errors without additive noise. In each panel, we vary one of the four properties of the calibration error while holding the others at set values, as listed in the caption. The scenarios investigated here are realistic in the calibrated data products in the existing observing runs: $\fwidth \sim 10$--$100$~Hz, $|\dApeak| \lesssim 10\%$, and $|\delta\phi_{\rm peak}| \lesssim 10^\circ$~\cite{Sun_2020,sun2021characterization}. The black horizontal lines indicate $p(M_t, \chi_t) = 0$ from correctly calibrated data. As expected, larger $|\dApeak|$, larger $|\dphipeak|$, and calibration errors peaked around the QNM frequencies lead to a more significant bias in the estimates of BH mass and spin. In panel (a), where we plot $p(M_t, \chi_t)$ as a function of $\Delta f$ (i.e., the frequency offset between the error peak and the 220 mode), the results are less symmetric when the peak calibration error moves away from the QNM frequency towards lower frequencies. The largest impact occurs slightly to the left of the 220 QNM frequency. Such asymmetry can partly be explained by the superposition of the positive and negative frequency components (see e.g., Ref.~\cite{Li2022,Giesler2025}), where the joint effect is affected by a phase factor determined by the source system. The effect of this superposition mainly occurs at lower frequencies, where the positive and negative components are closer to each other in frequency. The results at lower frequencies in panel (a) also show less smooth features than at higher frequencies, likely due to additional mode components and numerical noise in the NR waveform; see Appendix~\ref{app:manual} for further discussion.

Similarly, Fig.~\ref{fig:pval330} shows results for the SXS:BBH:0181 system (injecting ${}_{-2}Y_{2\pm2}$ and ${}_{-2}Y_{3\pm3}$ components only) by using a three-mode filter for $\{220, 221, 330\}$ QNMs. The results are qualitatively similar to those in Fig.~\ref{fig:pval0305}. The black horizontal lines mark $p(M_t, \chi_t) = 0.01$ from correctly calibrated data, deviating slightly from zero, likely due to overtones of the $\ell=m=3$ mode (as demonstrated in Appendix~\ref{app:manual}). In panel (a),  we find a secondary peak effect around the 330 QNM frequency at 322.54~Hz. The most significant bias still occurs to the left of the 220 QNM frequency.
In panel (b), we evaluate the calibration error's impact by varying $f_{\text{width}}$ under two scenarios: $f_{\rm peak} = f_{220}$ and $f_{\rm peak} = f_{330}$.
The largest impact occurs at $f_{\rm width} \sim 60$~Hz, slightly wider than the case in Fig.~\ref{fig:pval0305} (peaking at $f_{\rm width} \sim 50$~Hz).
This is because when there are both 220 and 330 in the signal, a slightly broader calibration error impacts both the 220 and 330 QNMs. However, since 220 is the dominant mode, calibration errors more closely aligned with the 220 mode always affect the results more significantly. Generally, a miscalibration impacts the parameter estimation in ringdown the most when we have $f_\text{peak}\sim f_{220}$, $\fwidth\sim 50$ Hz, a negative $\dApeak$, and a positive $\dphipeak$.

\subsection{Uncertainties caused by noise}
\label{sec:white_noise}
The presence of noise introduces additional uncertainty. Next, we quantify the impact of calibration errors on QNM analysis by simulating signals in additive noise. We repeat the analysis as shown in Fig.~\ref{fig:pval0305} (a) (no additive noise) by injecting the waveforms in white Gaussian noise. Given that the 220 and 221 QNMs we study here are at frequencies around 250~Hz, where the noise amplitude spectral density (ASD) can be approximately treated as flat. (More realistic, colored noise simulations are presented in later sections.) We inject the signal in $N=100$ random realizations of white Gaussian noise with a fixed ASD that produces a fiducial ringdown $\text{SNR}\sim 60$. The resulting $p(M_t, \chi_t)$ values are expected to be uniformly distributed given the correct signal hypothesis in the accurately calibrated data~\cite{talts2020}, with a mean of $\sim 0.5$ and a standard deviation of $\sim 0.29$. 
Miscalibration may lead to deviations in the joint posteriors of BH parameters, as shown in Fig.~\ref{fig:no_noise_contour}, which in turn, results in higher $p(M_t, \chi_t)$ values in general and a non-uniform distribution.

\begin{figure}
    \includegraphics[width=\columnwidth]{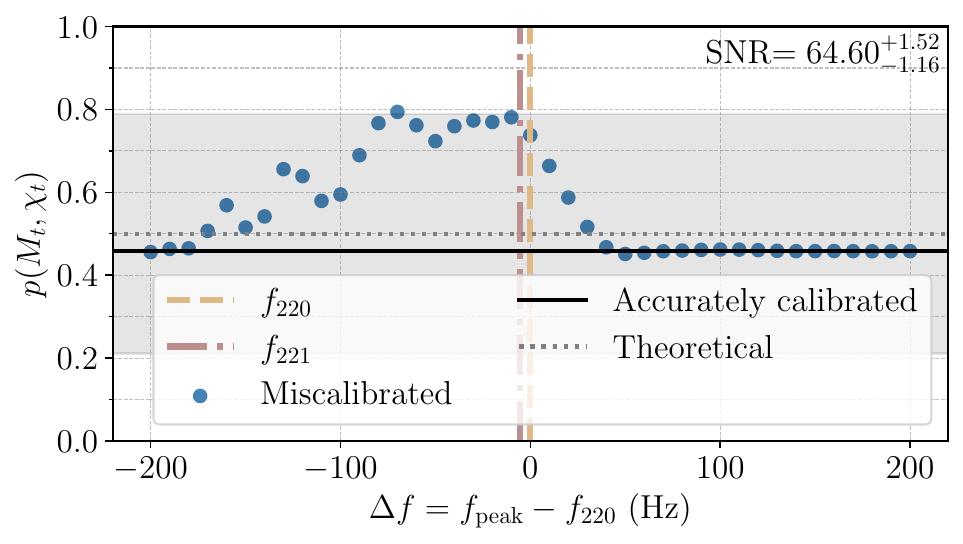}
    \caption{\label{fig:gausscentre} Joint posterior quantile values, $p(M_t, \chi_t)$, for a GW150914-like NR waveform (SXS:BBH:0305, ${}_{-2}Y_{2\pm2}$ components only) injected into white Gaussian noise, recovered with a $\{220,221\}$ filter and plotted as a function of the offset between the peak frequency of the calibration error and the 220 QNM frequency, $\Delta f = f_{\text{peak}}-f_{220}$ (other error parameters: $f_{\text{width}}=50$ Hz, $\dApeak=-10\%$, $\dphipeak = 10^\circ$). 
    Each blue dot is the mean $p(M_t, \chi_t)$ from $100$ noise realizations of the miscalibrated data (for a fixed error). The black line indicates the mean $p(M_t, \chi_t)$ from 100 noise realizations of accurately calibrated data. The gray dotted line and shade indicate the theoretical mean and $\pm 1\sigma$ bounds, respectively, for a uniform distribution.}
\end{figure}
\begin{figure}
    \includegraphics[width=\columnwidth]{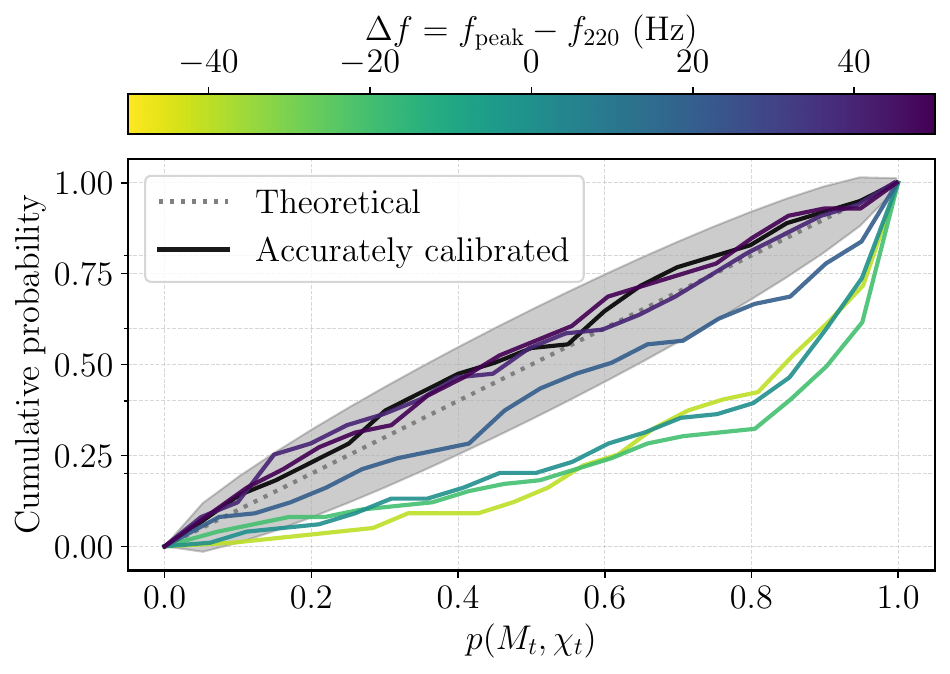}
    \caption{\label{fig:white_noise_pp_centre}Cumulative probability distributions of the joint posterior quantile, $p(M_t, \chi_t)$, for a GW150914-like NR waveform (SXS:BBH:0305, ${}_{-2}Y_{2\pm2}$ components only) injected into white Gaussian noise, recovered with a $\{220,221\}$ filter. Colored curves are results obtained from data sets containing calibration errors peaked at different frequencies, with the color indicating $\Delta f = f_{\text{peak}}-f_{220}$ (other error parameters: $f_{\text{width}}=50$ Hz, $\dApeak=-10\%$, $\dphipeak = 10^\circ$). The distribution of $p(M_t, \chi_t)$ obtained from the accurately calibrated data is shown in black. The dotted diagonal and shaded region indicate the theoretically expected distribution with 3$\sigma$ bounds.}
\end{figure}

Fig.~\ref{fig:gausscentre} shows the resulting $p(M_t, \chi_t)$ as a function of $\Delta f = \fpeak-f_{220}$. The other calibration error parameters are fixed at $f_{\text{width}}=50$ Hz, $\dApeak=-10\%$, and $\dphipeak=10^\circ$. The blue dots and black line are obtained from miscalibrated data and accurately calibrated data, respectively. The gray line and shade represent the expected theoretical mean and standard deviation for a uniform distribution. The total SNR of the full ringdown wave in each noise realization also differs. We note the median and $\pm 1 \sigma$ of the SNRs from all the trials in the top right corner of the plot. In the presence of noise, it is expected that the accurately calibrated data do not yield a perfect recovery of $(M_t, \chi_t)$, that is, the black line does not align with zero. However, the mean value of $p(M_t, \chi_t)$ from the accurately calibrated data does agree with the theoretical mean (i.e. close to the gray dotted line and within the gray shade). When the calibration errors peak at frequencies slightly lower than the 220 mode frequency, the bias caused by miscalibration is most significant, with the mean (blue dots) lying near the $1\sigma$ theoretical bound (gray shade). When $\fpeak$ moves away from $f_{220}$ towards higher frequencies, e.g., $\Delta f >50$ Hz, there is negligible impact on the analysis from the miscalibration; the results from the miscalibrated data largely align with those obtained from the data without calibration errors, with a maximum difference of 0.03 in the mean $p(M_t, \chi_t)$ values. The trend matches the results in Fig.~\ref{fig:pval0305} (a), when there is no additive noise. 

Fig.~\ref{fig:white_noise_pp_centre} shows the cumulative probability distributions of $p(M_t, \chi_t)$ for selected error scenarios in Fig.~\ref{fig:gausscentre} with fixed $f_{\text{width}}$, $\dApeak$, and $\dphipeak$, but varying $f_{\rm peak}$, as indicated by the color of the curves. The distribution obtained from accurately calibrated data is shown in black. The diagonal and shaded gray region indicate the theoretically predicted uniform distribution of $p(M_t, \chi_t)$ and the associated $3\sigma$ bounds. 
Fig.~\ref{fig:white_noise_pp_centre} presents the results shown in Fig.~\ref{fig:gausscentre} in an alternative way and demonstrates that as $\fpeak$ moves away from $f_{220}$ toward lower frequencies, the results are biased, and the distributions moves outside the $3\sigma$ bounds. Without prior knowledge of calibration errors, such a deviation from theoretically predicted uniform distribution of $p(M_t, \chi_t)$ might be misinterpreted as evidence that the ringdown model is incorrect. This could lead to the erroneous inclusion of alternative or additional modes in an attempt to improve the results~\cite{Lu2025}.
For example, in simulations where a calibration error overlaps with the 220-mode frequency, we find that incorrectly including a second overtone (222) can occasionally spuriously improve the agreement with the true BH parameters $(M_t, \chi_t)$.
In addition, such deviations might also be misinterpreted as evidence of deviations from GR, further complicating the interpretation of the results.

\section{Calibration accuracy requirement}
\label{sec:requirements}
In this section, we investigate the calibration accuracy required in order to perform robust ringdown analyses in colored detector noise. Secs.~\ref{sec:ligo} and \ref{sec:ce} focus on the current- and next-generation detectors, respectively.

\subsection{\label{sec:ligo}Current generation detectors}

Given that the current state-of-the-art calibration accuracy is $|\delta \mathcal{A}|\lesssim 10\%$ and $|\delta \phi| \lesssim 10^\circ$ in aLIGO \cite{Sun_2020,sun2021characterization}, we first investigate whether the current level of calibration error impacts the ringdown analyses with current-generation detectors. 

Since the overall trends of the impacts from calibration errors are expected to be similar to the results we obtained in Sec.~\ref{sec:method}, we fix $f_{\text{peak}}=f_{220}$ and $f_{\text{width}}=50$~Hz for the rest of the study, which generally leads to the worst bias in the analysis, as shown in Fig.~\ref{fig:pval0305} [panels (a) and (b)], and vary $\dApeak$ and $\dphipeak$. 
At higher dimensions, different error characteristics can exhibit degeneracies; for instance, correlated variations in $f_{\text{peak}}$ and $f_{\text{width}}$ may result in similar effects. A detailed investigation of these more complex simulated scenarios, along with studies incorporating real calibration errors, is deferred to future work.

At the aLIGO design sensitivity~\cite{asdcurves}, we use the aLIGO Hanford as an example for detector location and test the GW150914-like NR waveforms (SXS:BBH:0305) at a luminosity distance of $D_L=410$ Mpc (similar to the real GW150914 event) \cite{Abbott_2016propsim}. Here, we adopt both the $p(M_t, \chi_t)$ distribution and the distance between the MAP estimates and the true BH parameters, $\epsilon$, defined in Eq.~\eqref{eq:epsilon}, to provide a comprehensive evaluation. Note that $p(M_t, \chi_t)$ distributions deviating more from the theoretically predicted uniform distribution, corresponding to a larger mean of $p(M_t, \chi_t)$ above 0.5, indicate worse recovery of the BH parameters, and so do larger values of $\epsilon$.

\begin{figure*}[tbh]
    \includegraphics[width=0.9\linewidth]{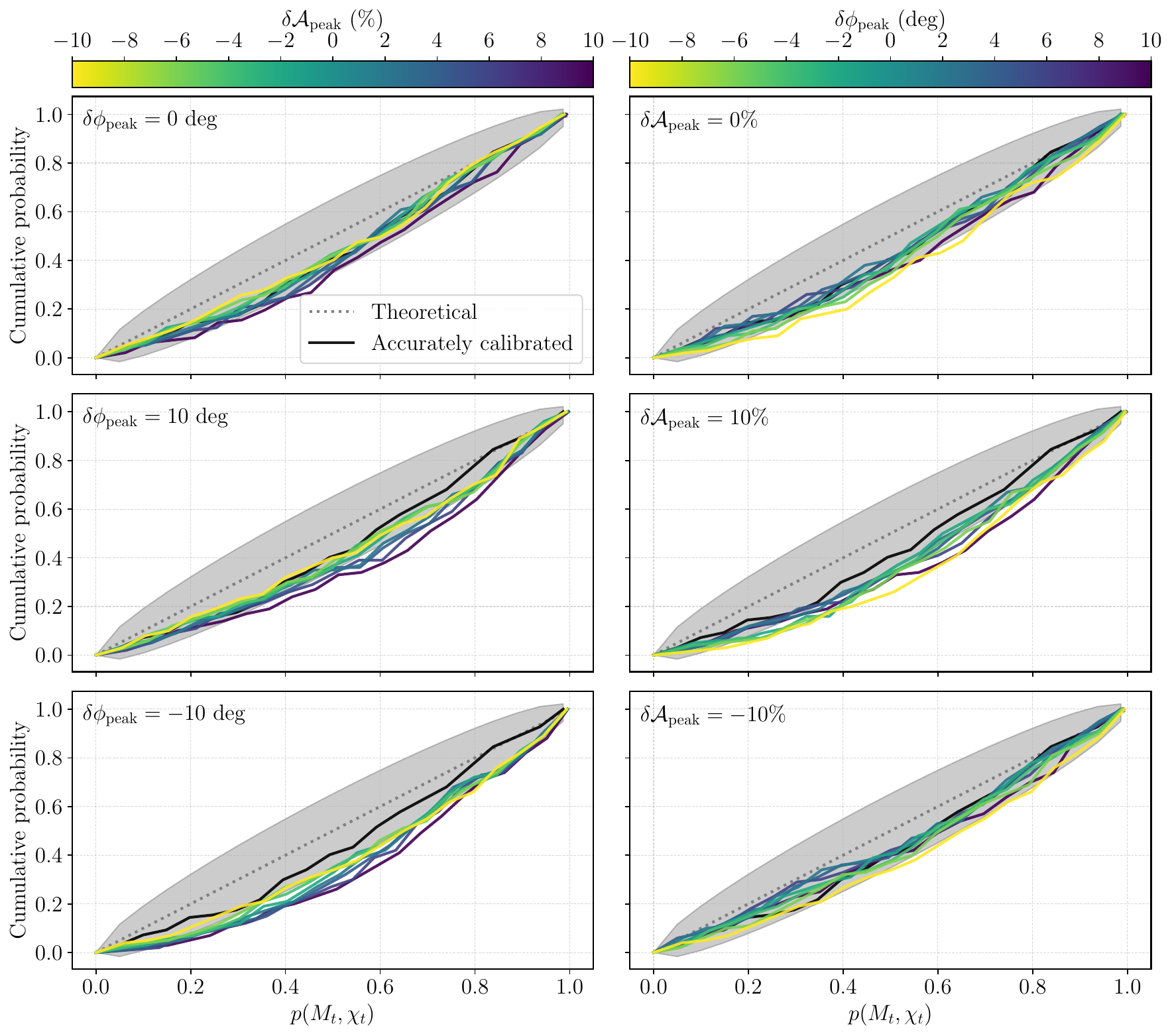}\caption{\label{fig:0305_h1_p}Cumulative probability distributions of the joint posterior quantile, $p(M_t, \chi_t)$, for a GW150914-like NR waveform (SXS:BBH:0305, ${}_{-2}Y_{2\pm2}$ components only) injected into aLIGO Hanford at the design sensitivity, recovered with a $\{220,221\}$ filter ($D_L = 410$ Mpc, equivalent to a total ringdown SNR of $\sim 33$). 
    Left: Colored curves show results from miscalibrated data sets containing errors with different $\dApeak$ (color bar) and three choices of fixed $\dphipeak$ (noted on each subplot).
    Right: Colored curves show results from miscalibrated data sets containing errors with different $\dphipeak$ (color bar) and three choices of fixed $\dApeak$ (noted on each subplot).
    Other error parameters are fixed at $\fpeak=f_{220}$ and $\fwidth=50$ Hz. The distribution of $p(M_t, \chi_t)$ obtained from the accurately calibrated data is shown in black. The dotted diagonal and shaded region indicate the theoretically expected uniform distribution with 3$\sigma$ bounds.
    }
\end{figure*}
\begin{figure*}
    \includegraphics[width=0.9\linewidth]{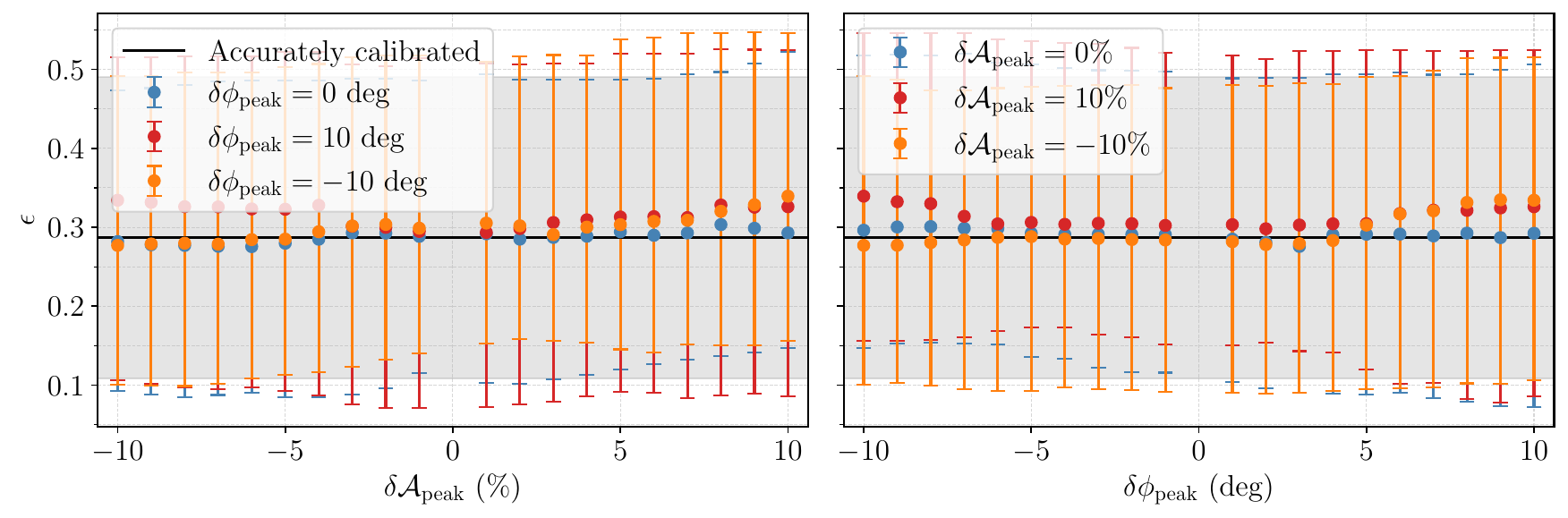}\caption{\label{fig:0305_h1_dist}Parameter distance, $\epsilon$, for a GW150914-like NR waveform (SXS:BBH:0305, ${}_{-2}Y_{2\pm2}$ components only) injected into aLIGO Hanford at the design sensitivity, recovered with a $\{220,221\}$ filter ($D_L = 410$ Mpc, equivalent to a total ringdown SNR of $\sim 33$). 
    Left: $\epsilon$ versus varying peak magnitude error ($\dApeak$) with three choices of fixed peak phase error ($\dphipeak$). 
    Right: $\epsilon$ versus varying peak phase error ($\dphipeak$) with three choices of fixed peak magnitude error ($\dApeak$). 
    Other error parameters are fixed at $\fpeak=f_{220}$ and $\fwidth=50$ Hz. The markers with error bars are the median and $\pm 1\sigma$ bounds from the miscalibrated data (see legend). The black line with shade represents the median results from the accurately calibrated data with the $\pm 1\sigma$ bounds. 
    }
\end{figure*}

Figs.~\ref{fig:0305_h1_p} and~\ref{fig:0305_h1_dist} show the results. In Fig.~\ref{fig:0305_h1_p}, the $p(M_t, \chi_t)$ distribution from accurately calibrated data (black curve) deviates slightly from the theoretical prediction (dotted diagonal) but remains within the 3$\sigma$ bounds, due to the limited total ringdown SNR of $\sim 33$ (cf. Fig.~\ref{fig:white_noise_pp_centre}, obtained at a ringdown SNR of $\sim 65$).
In Fig.~\ref{fig:0305_h1_dist}, we plot the parameter distance, $\varepsilon$, which exhibits a trend consistent with the mean value of $p(M_t, \chi_t)$.\footnote{In simulations without additive noise, $\varepsilon$ exhibits the same behavior as $p(M_t, \chi_t)$. Since the QNMs considered here lie within the most sensitive frequency band of the detectors (hundreds of Hz), the results are largely dominated by the SNR, with negligible differences between simulations with white Gaussian noise and those with colored detector noise at the design sensitivity. For clarity, we therefore present in this section only the $\varepsilon$ results obtained with colored noise simulations.} Among all combinations of different choices of $\dApeak$ and $\dphipeak$, the results obtained from miscalibrated data and accurately calibrated data are generally consistent. The most extreme calibration errors (i.e., $|\dApeak|\sim 10\%$ and $|\dphipeak|\sim 10^\circ$) produce a slightly biased distribution, pushing the $p(M_t, \chi_t)$ distribution out of the 3-$\sigma$ bounds of the theoretical prediction. The impact on the $\epsilon$ distribution is minor, albeit present, with the largest difference between the median values obtained from miscalibrated and accurately calibrated data being 0.06. Larger calibration errors generally lead to slightly worse parameter recovery in both $p(M_t, \chi_t)$ and $\epsilon$. We expect that similar conclusions can be obtained for higher-order angular modes, such as a 330 mode. As demonstrated in Figs.~\ref{fig:pval0305} and~\ref{fig:pval330}, the bias introduced by miscalibration is on the same level for the two sets of waveforms. As such, we do not repeat the tests shown in Figs.~\ref{fig:0305_h1_p} and \ref{fig:0305_h1_dist} for the high-mass-ratio waveform, SXS:BBH:0181. 

Overall, the current standard of calibration errors, $|\delta \mathcal{A}|\lesssim 10\%$ and $|\delta \phi| \lesssim 10^\circ$, does not affect the ringdown analysis results in the relatively low-SNR regime (ringdown SNR $\lesssim 30$)\footnote{We consider the low-SNR regime relevant to this study--specifically, the minimum SNR at which calibration errors begin to affect the analysis--whereas in typical GW analyses, SNRs $\lesssim 10$ are generally regarded as low.} 
with current-generation detectors when the uncertainties in parameter estimation are dominated by noise, except for some worst-case scenarios where $|\delta \mathcal{A}|\sim 10\%$, $|\delta \phi| \sim 10^\circ$, and the error happens to align with the dominant QNM in frequency domain.
While such a worst-case error scenario is likely rare in existing observations~\cite{Sun_2020, sun2021characterization}, the analysis results may lead to misinterpretation of the accuracy of the ringdown model and the consistency with GR, if such calibration systematics are unaccounted for.

\begin{figure*}[!tbh]
    \includegraphics[width=0.92\linewidth]{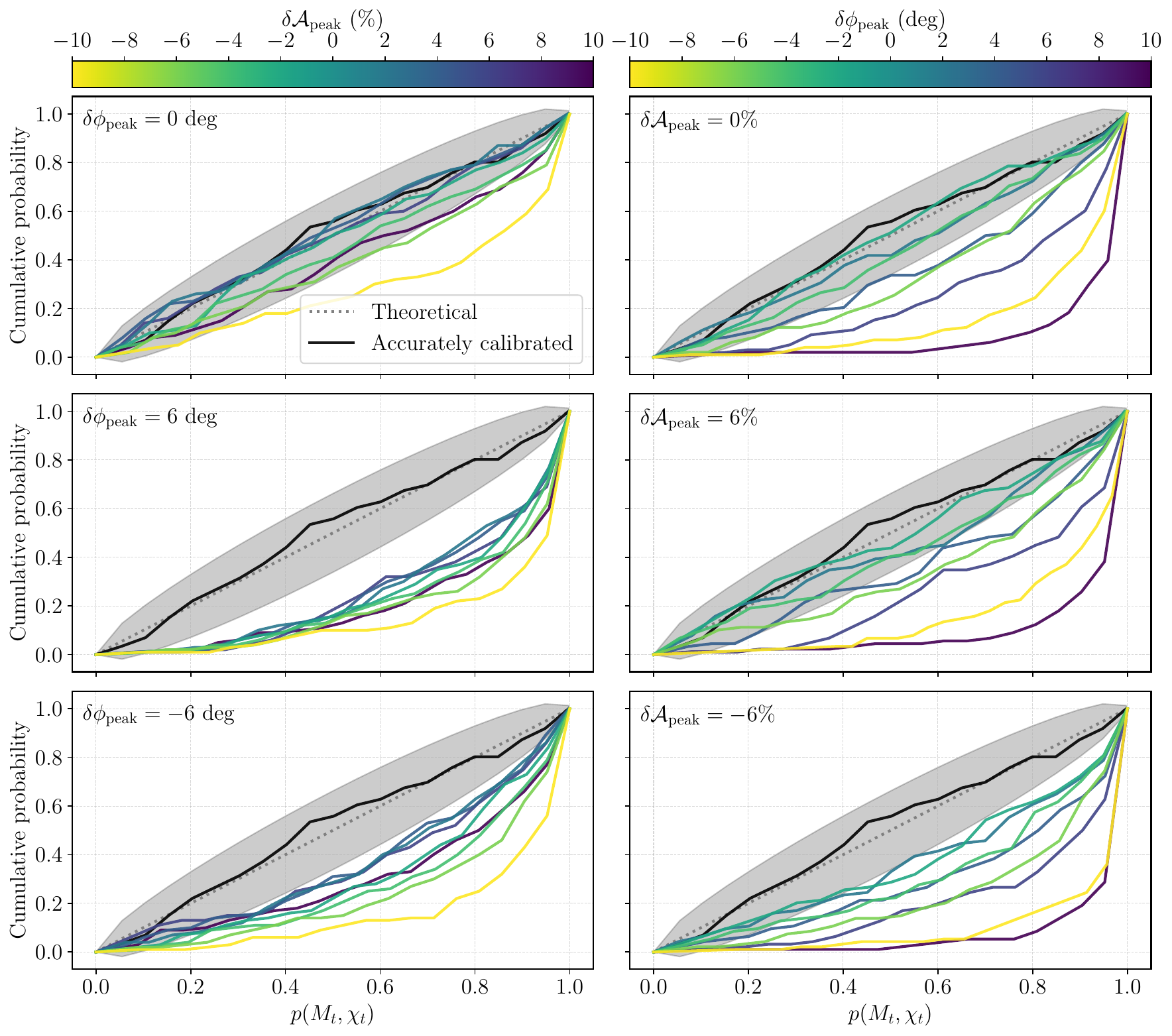}
    \caption{\label{fig:0305_CE_p}(Similar to Fig.~\ref{fig:0305_h1_p}) Cumulative probability distributions of the joint posterior quantile, $p(M_t, \chi_t)$, for a GW150914-like NR waveform (SXS:BBH:0305, ${}_{-2}Y_{2\pm2}$ components only) injected into Cosmic Explorer, recovered with a $\{220,221\}$ filter ($D_L = 2$ Gpc, equivalent to a total ringdown SNR of $\sim 125$).
    }
\end{figure*}

\begin{figure*}[!tbh]
    \includegraphics[width=0.92\linewidth]{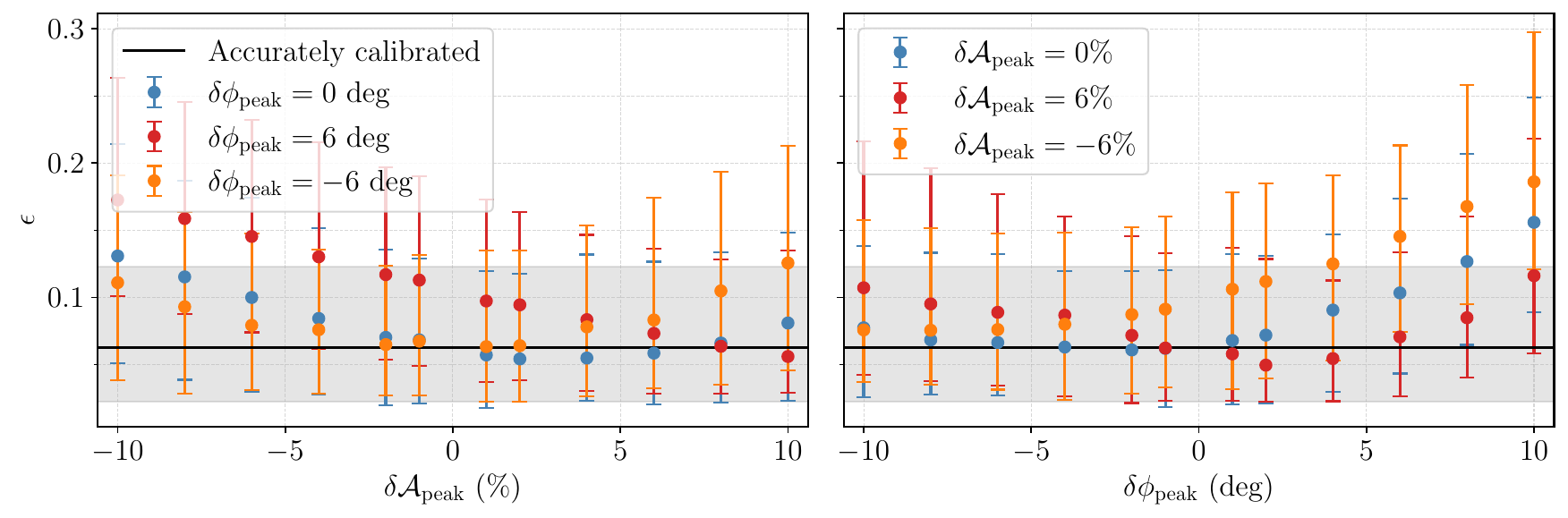}
    \caption{\label{fig:0305_CE_dist}(Similar to Fig.~\ref{fig:0305_h1_dist}) Parameter distance, $\epsilon$, for a GW150914-like NR waveform (SXS:BBH:0305, ${}_{-2}Y_{2\pm2}$ components only) injected into Cosmic Explorer, recovered with a $\{220,221\}$ filter ($D_L = 2$ Gpc, equivalent to a total ringdown SNR of $\sim 125$).
    }
\end{figure*}

\subsection{\label{sec:ce} Next generation detectors}

With next-generation observatories, e.g., Cosmic Explorer (CE)~\cite{evans2021horizon} and Einstein Telescope (ET)~\cite{Maggiore2020}, we will be able to do BH spectroscopy with high-SNR events (ringdown SNR $\gtrsim 120$), revealing more subdominant modes and carrying out precision tests of GR. In this section, we evaluate the impact of calibration errors by simulating high-SNR signals in CE~\cite{evans2021horizon, evans2023cosmic, CEasd2022}. We test the GW150914-like (SXS:BBH:0305) and the high-mass-ratio (SXS:BBH:0181) systems at a ringdown SNR $\sim 125$ (corresponding to $D_L=2$ Gpc and 600 Mpc, respectively). In the case of extremely high-SNR (ringdown SNR $\gtrsim 250$) events, e.g., a GW150914-like system at $D_L=410$ Mpc, which yields a total ringdown SNR of $\sim 600$ in CE, parameter estimations would require a much higher resolution in the parameter space, and more careful scrutiny is required on other potential impacts from, e.g., waveform systematics. We defer a comprehensive study of the extremely high SNR regime to future work. 

\begin{figure*}[!tbh]
    \includegraphics[width=0.92\linewidth]{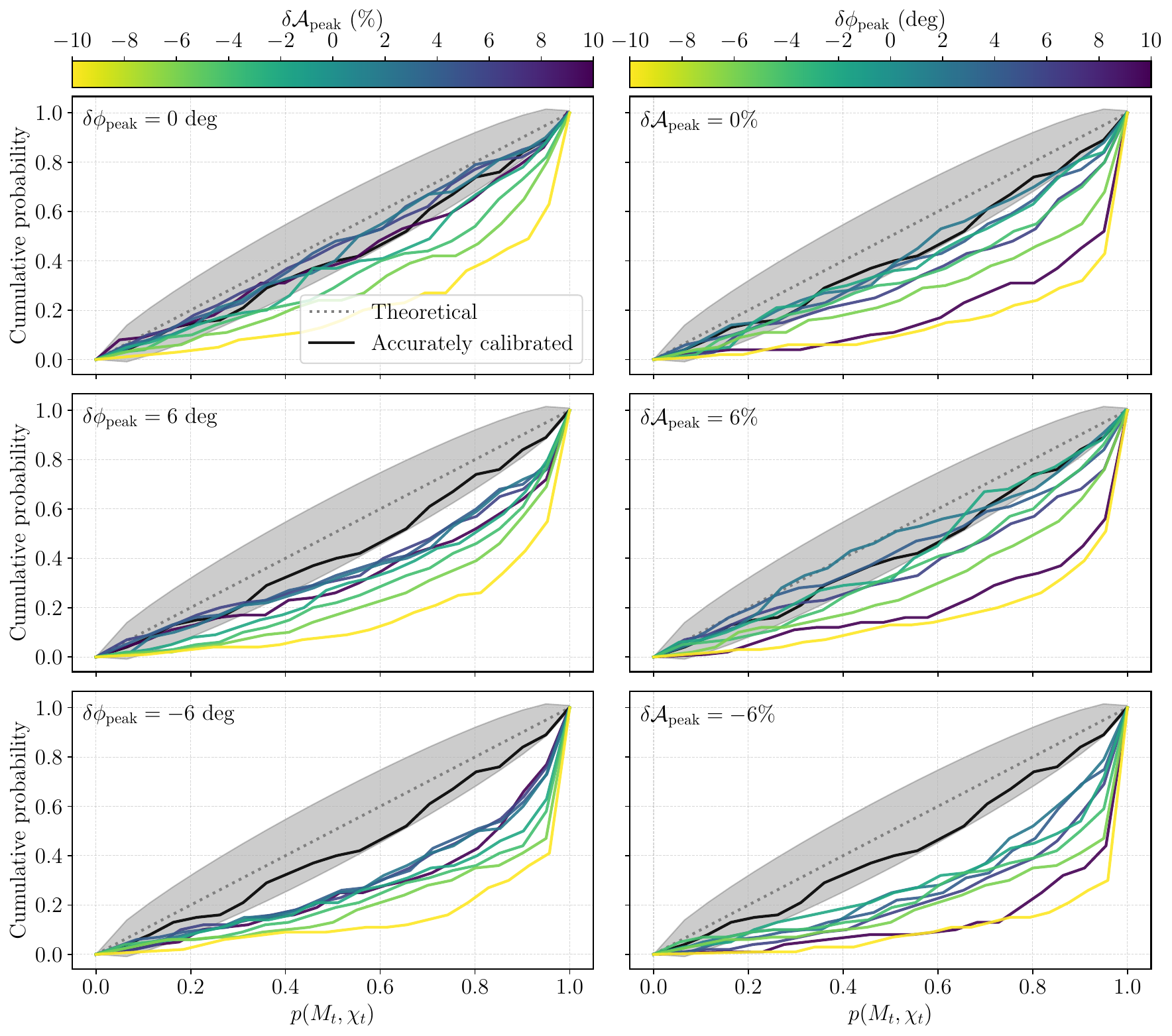}
    \caption{\label{fig:0181_CE_p}(Similar to Fig.~\ref{fig:0305_h1_p}) Cumulative probability distributions of the joint posterior quantile, $p(M_t, \chi_t)$, for a high-mass-ratio NR waveform (SXS:BBH:0181, ${}_{-2}Y_{2\pm2}$ and ${}_{-2}Y_{3\pm3}$ components) injected into Cosmic Explorer, recovered with a $\{220,221, 330\}$ filter ($D_L = 600$ Mpc, equivalent to a total ringdown SNR of $\sim 126$).
    }
\end{figure*}

\begin{figure*}[!tbh]
    \includegraphics[width=0.92\linewidth]{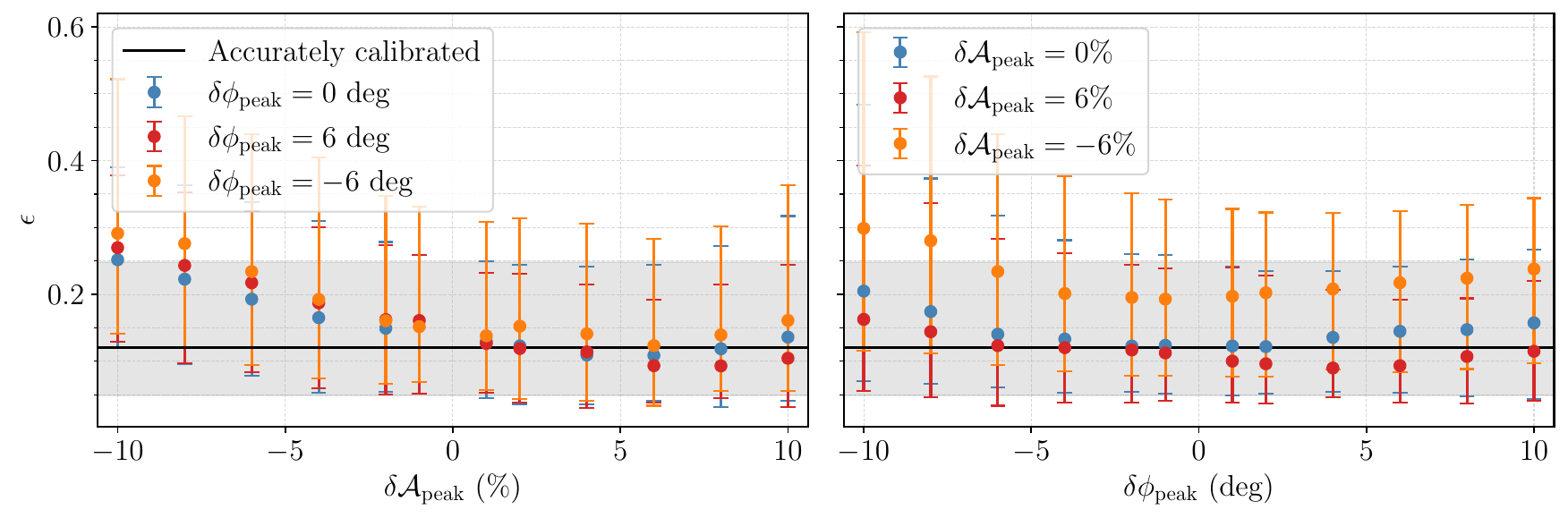}
    \caption{\label{fig:0181_CE_dist}(Similar to Fig.~\ref{fig:0305_h1_dist}) Parameter distance, $\epsilon$, for a high-mass-ratio NR waveform (SXS:BBH:0181, ${}_{-2}Y_{2\pm2}$ and ${}_{-2}Y_{3\pm3}$ components) injected into Cosmic Explorer, recovered with a $\{220,221, 330\}$ filter ($D_L = 600$ Mpc, equivalent to a total ringdown SNR of $\sim 126$).
    }
\end{figure*}

Figs.~\ref{fig:0305_CE_p}--\ref{fig:0181_CE_dist} show simulation results of SXS:BBH:0305 and SXS:BBH:0181 in CE, with the same scenarios as presented in Figs.~\ref{fig:0305_h1_p} and~\ref{fig:0305_h1_dist}.
The BH parameter recovery in accurately calibrated data is improved in CE due to the higher SNR compared to that in aLIGO. The $p(M_t, \chi_t)$ distribution from accurately calibrated data sets in Fig.~\ref{fig:0305_CE_p} agrees with the theoretical prediction to a greater degree compared to Fig.~\ref{fig:0305_h1_p}. As shown in Figs.~\ref{fig:0305_h1_dist} and \ref{fig:0305_CE_dist}, the median $\epsilon$ decreases to $\sim 0.06$ in CE (c.f. median $\epsilon \sim 0.28$ in aLIGO). 
Comparing the two waveforms tested with CE sensitivity, the $p(M_t, \chi_t)$ distribution and $\epsilon$ in accurately calibrated data is slightly better for the signal consisting of the $\ell=m=2$ QNMs only. This is as expected (similar to the study for pure signal waveform in Figs.~\ref{fig:pval0305} and \ref{fig:pval330}) due to the fact that additional overtones and mode mixing effects come into play when including higher-order angular modes. In order to achieve more accurate parameter estimation for the SXS:BBH:0181 waveform, a more complex ringdown model needs to be investigated (e.g., by including more QNMs). 
To evaluate the impact of calibration errors, we compare results from miscalibrated data to those from accurately calibrated data, and the slight imperfection in the ringdown model does not affect our findings. 
Thus, in this study, we do not explore more complex ringdown models for SXS:BBH:0181.\footnote{Note that when a filter, $\mathcal{F}_{\ell mn}$, is applied to the data, the amplitude of other QNMs is suppressed by a factor $B_{\ell mn}^{\ell 'm'n'}$ (see Sec.~\ref{sec:qnm}). Thus, a higher total ringdown SNR is required for similar analysis when adding more modes to the ringdown model. A different analysis starting time may also be required (see Appendix~\ref{app:time}).}


Figs.~\ref{fig:0305_CE_p} and \ref{fig:0181_CE_p} demonstrate calibration errors with $|\delta \mathcal{A}| \sim 10\%$ and/or $|\delta \phi| \sim 10^\circ$ can lead to significant bias in parameter estimation in the next-generation detectors, pushing the $p(M_t, \chi_t)$ distributions completely outside the theoretical $3\sigma$ bounds. Such biases could lead to the misinterpretation of the QNM content present in the ringdown signal, or even erroneously indicate a deviation from GR. This is also reflected in Figs.~\ref{fig:0305_CE_dist} and \ref{fig:0181_CE_dist}, where the median $\epsilon$ values from the miscalibrated data lie outside the $1\sigma$ bounds obtained in the accurately calibrated data. The most significant biases seen among the scenarios tested occur with a big phase error of $|\dphipeak| = 10^\circ$. Generally, a calibration accuracy of $|\dApeak| \lesssim 4\%$ and $|\dphipeak| \lesssim 4^\circ$ is required to keep estimated parameters generally consistent with those from accurately calibrated data to the $3\sigma$ level. Thus, the state-of-the-art calibration accuracy in the existing observing runs with current-generation observatories may not satisfy the requirements for precision tests of GR and BH spectroscopy with next-generation observatories. More accurate calibration is required for future detectors.

\section{\label{sec:conclusion} Conclusion}

We use the QNM rational filter---a method for analyzing ringdown signals---to evaluate the impact of calibration systematics on BH spectroscopy and tests of GR. We use a set of physically motivated, parametrized, synthetic calibration errors to `miscalibrate' NR waveforms and compare the ringdown analysis results in miscalibrated data to those in accurately calibrated data with a variety of noise scenarios. We confirm that the current calibration accuracy of magnitude errors $\lesssim 10\%$ and phase errors $\lesssim 10^\circ$ in current ground-based GW observatories is, barring extreme scenarios, sufficient for tests of GR, but find that in future-generation observatories, calibration errors with peak magnitude error $\gtrsim 4\%$ or peak phase error $ \gtrsim 4^\circ$ can lead to significantly biased results. This is primarily because future observatories will achieve higher SNRs. In particular, extreme SNR cases (e.g., SNR $\gtrsim 250$) will likely demand even more stringent calibration requirements. 

Observations of GWs from CBCs enable tests of GR in the strong-field regime through a variety of methods. Currently, inconclusive or ambiguous results can often be attributed to the noise fluctuations in the low SNR regime (ringdown SNR $\lesssim 30$). However, this will not hold true in the future, as increasing SNRs will reduce statistical uncertainties. In such scenarios, unaccounted-for systematics could obscure genuine signal features or falsely appear as violations of GR. Indeed, we find this to be the case for BH spectroscopy at sufficiently high SNRs (ringdown SNR $\gtrsim 120$). This study finds that as we move into a higher-SNR regime with current detectors, it is essential to take into account these systematics explicitly. In particular, while analyses such as CBC parameter estimation of the full inspiral-merger-ringdown signal already incorporate estimated calibration systematics, the omission in other tests, e.g., analyses focusing on ringdown waves, could introduce inconsistencies between the results derived from the full waveform and those from focused analyses. As we enter the design phase for the next-generation detectors, which demands the development of new calibration technologies~\cite{evans2021horizon}, it may be necessary to establish minimum calibration requirements. Overall, we conclude that as the sensitivity of GW detectors increases, testing GR analyses must account for calibration systematics to avoid biased or misleading results. Consequently, next-generation detectors will require more stringent calibration standards to conduct robust and conclusive tests of GR. 

The results obtained in this study have certain limitations. We have investigated only a small subset of systems, with limited combinations of the calibration magnitude and phase errors, and focusing on a moderate SNR regime. Different binary merger systems can excite a range of different modes during the ringdown phase, extending beyond just the $\ell=m=2$ and $\ell=m=3$ modes. Furthermore, next-generation GW observatories will observe events at extremely high SNRs (with ringdown SNRs reaching $\sim 600$). At such high SNRs, it is likely that far more stringent calibration accuracy will be necessary for robust BH spectroscopy. A comprehensive investigation with more complex mode excitations in combination with more complex calibration systematics (including simulations with real calibration errors) will be conducted as future work.

\begin{acknowledgments}
The authors thank Neil Lu and Ornella Piccinni for valuable discussions, and Fiona Panther for her helpful comments during the internal review. This material is based upon work supported by NSF's LIGO Laboratory which is a major facility fully funded by the National Science Foundation. The authors are grateful for the computational resources provided by the OzSTAR national facility at Swinburne University of Technology. The OzSTAR program receives funding in part from the Astronomy National Collaborative Research Infrastructure Strategy (NCRIS) allocation provided by the Australian Government, and from the Victorian Higher Education
State Investment Fund (VHESIF) provided by the Victorian Government.
This research is supported by the Australian Research Council Centre of Excellence for Gravitational Wave Discovery (OzGrav), Project Numbers CE170100004 and CE230100016. LS is also supported by the Australian Research Council Discovery Early Career Researcher Award, Project Number DE240100206. Research at Perimeter Institute is supported in part by the Government of Canada through the Department of Innovation, Science and Economic Development and by the Province of Ontario through the Ministry of Colleges and Universities. 
\end{acknowledgments}

\appendix
\section{\label{app:time}Analysis start time}
\begin{figure}
    \includegraphics[width=\columnwidth]{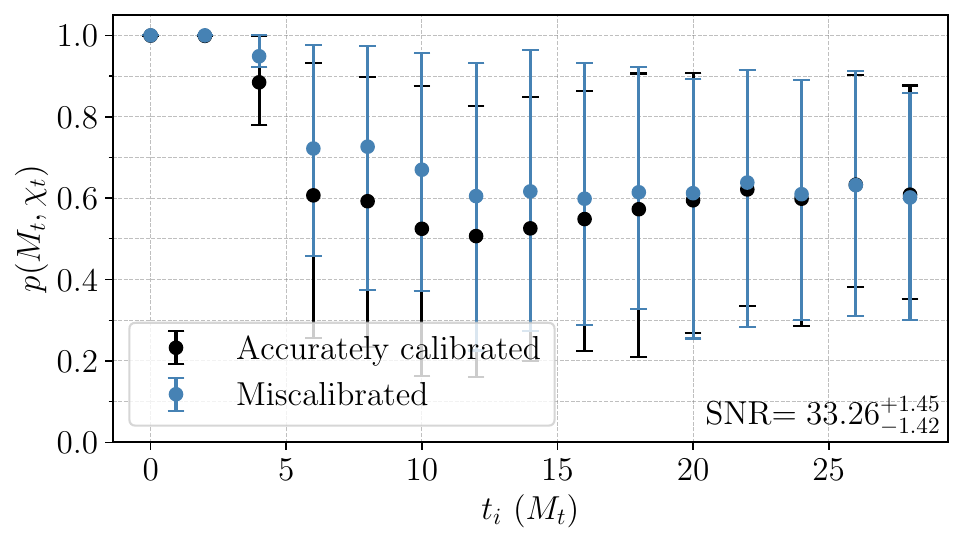}
    \caption{\label{fig:colH1time} Joint posterior quantile values, $p(M_t, \chi_t)$, for a GW150914-like NR waveform (${}_{-2}Y_{2\pm2}$ components only) injected into aLIGO Hanford at the design sensitivity, as a function of the analysis starting time, $t_i$. The signal is injected at a luminosity distance of $D_L = 410$ Mpc (similar to GW150914), equivalent to a total ringdown SNR of $\sim 33$ (with error bars around the SNRs indicating the 16th and 84th percentile among the 100 trials). The blue and black markers stand for the mean values from miscalibrated data and accurately calibrated data, respectively, with the error bars indicating the 16th and 84th percentiles across 100 trials. The calibration error parameters are $f_{\mathrm{peak}}=f_{220}$, $f_{\mathrm{width}}=50$ Hz, $\dApeak=-10\%$ and $\dphipeak = 10^\circ$. The $p(M_t, \chi_t)$ values show that the recovery of the BH properties is always worse in miscalibrated data for $4 M_t \lesssim t_i \lesssim 20 M_t$ when the 220 and 221 are both present.}
\end{figure}

Given that higher-order overtones decay quicker, when we focus on testing the 220 fundamental mode and the first overtone 221 mode (using the SXS:BBH:0305 waveform), we can neither set $t_i$ too early when higher-order overtones $(n\geq 2)$ are still present nor too late when the 221 mode has sufficiently decayed. We select a suitable $t_i$ which yields expected recovery of $p(M_t, \chi_t)$ in accurately calibrated data with a $\{220,221\}$ filter. The blue and black dots in Fig.~\ref{fig:colH1time} show the mean joint posterior quantile values, $p(M_t, \chi_t)$, among $100$ noise realizations, for miscalibrated and accurately calibrated data, respectively, with the error bars indicating the 16th and 84th percentiles. 
The calibration error is fixed at $f_{\mathrm{peak}}=f_{220}$, $f_{\mathrm{width}}=50$ Hz, $\dApeak=-10\%$, and $\dphipeak = 10^\circ$. A suitable analysis start time of $t_i=16 M_t$ is chosen and used throughout this study. A similar test was conducted for the high-mass-ratio NR waveform (SXS:BBH:0181) with the additional $\ell=m=3$ modes included.

\section{\label{app:manual} Method validation with manually constructed signals}

Before we conduct analysis with NR waveforms, we validate the procedure using a set of clean, controlled injections that are free of impacts from additional mode components, higher-order overtones, numerical noise, etc. that can be present in NR waveforms. By comparing the results from the controlled and NR injections, we investigate the features unique to NR waveforms and confirm that no impact on our results is introduced by the additional features associated with NR waveforms. 
We perform two such injections, both using mode frequencies consistent with a GW150914-like remnant BH but with arbitrarily chosen amplitudes and phases. One injection includes the 220 and 221 QNMs only, and the other includes the 220 and 330 QNMs only.

\begin{figure*}
    \includegraphics[width=.9\linewidth]{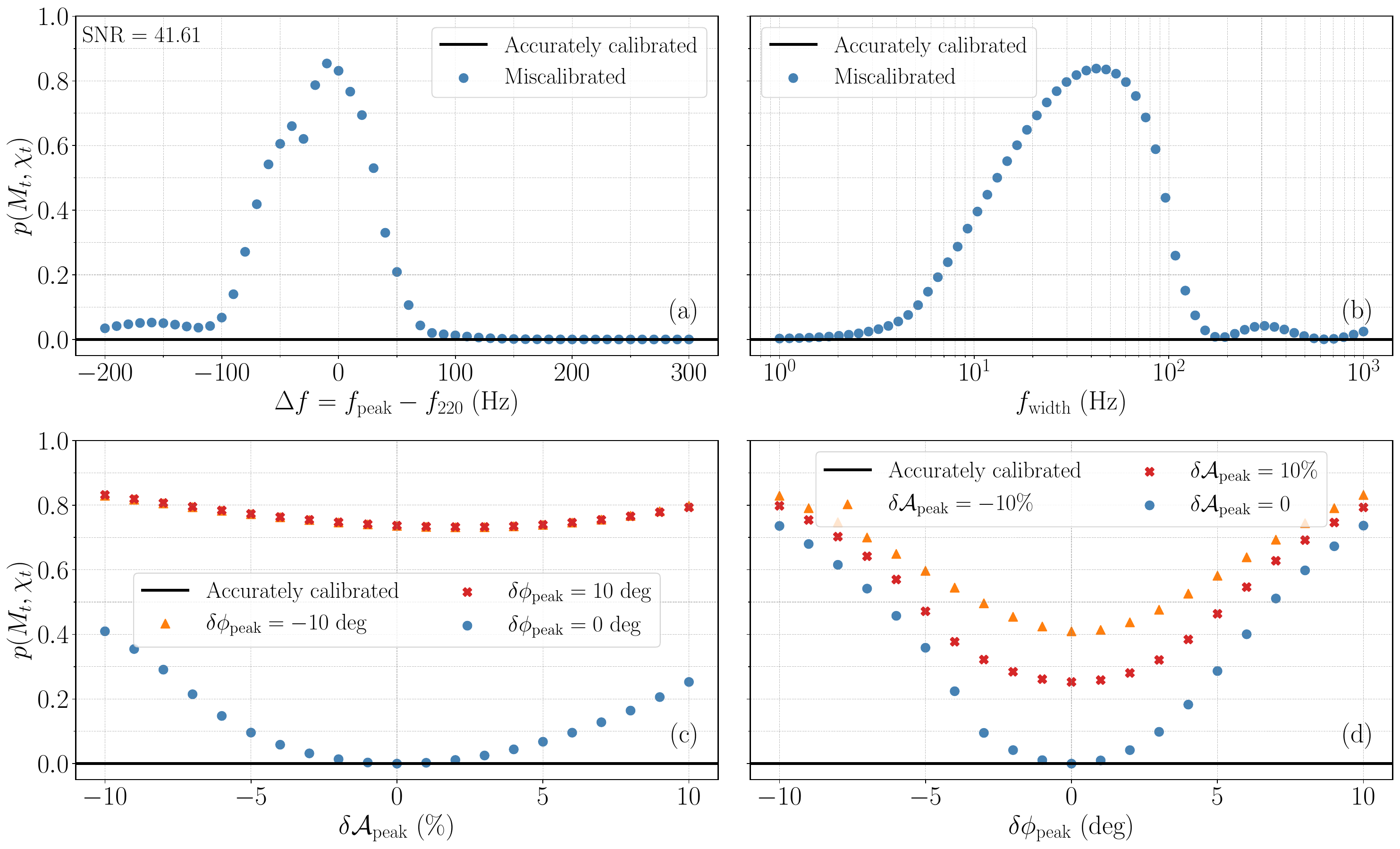}%
        \label{subfig:a220221}%

    \caption{\label{fig:nonoise220}(Similar to Fig.~\ref{fig:pval0305}; see details in Fig.~\ref{fig:pval0305} caption) Joint posterior quantile values, $p(M_t, \chi_t)$, for a manually constructed 220 and 221 signal from a GW150914-like system (not from an NR waveform) in aLIGO Hanford without additive noise as a function of the calibration error properties.
    }
\end{figure*}

Fig.~\ref{fig:nonoise220} shows a similar set of tests as presented in Fig.~\ref{fig:pval0305} for a manually-constructed, controlled 220 and 221 signal injected into aLIGO Hanford detector with various calibration errors without additive noise. The general trends between Figs.~\ref{fig:nonoise220} and~\ref{fig:pval0305} are similar. Overall, we see that larger $|\dApeak|$, larger $|\dphipeak|$ and calibration errors with $\fwidth\sim10$--$100$~Hz (comparable to the characteristic width of the QNM in the frequency domain) peaked around the QNM frequencies have greater impact on the estimates of BH mass and spin.\footnote{As a cross-check, we also look at the signals in both the time and frequency domains and compare the results with other studies to validate the method.} The differences between Figs.~\ref{fig:nonoise220} and \ref{fig:pval0305} come from additional mode components and the numerical noise in the NR waveform. The most obvious difference is in panel (a), which shows $p(M_t, \chi_t)$ as a function of $\Delta f$. For the NR waveform, the results are less symmetric when the calibration error peak moves away from the QNM frequency, partly explained by the superposition of the positive and negative frequency components, which involves a phase factor determined by the source system. The phase factor is different between the NR waveform and the manually constructed signal. In addition, minor differences in the trends between Figs.~\ref{fig:nonoise220} and \ref{fig:pval0305} may be due to the NR waveform containing higher-order overtones, retrograde modes, other angular modes mixed into ${}_{-2}Y_{2\pm2}$ directions, and numerical noise, which do not affect the results in our analysis. 

\begin{figure*}
        \includegraphics[width=.9\linewidth]{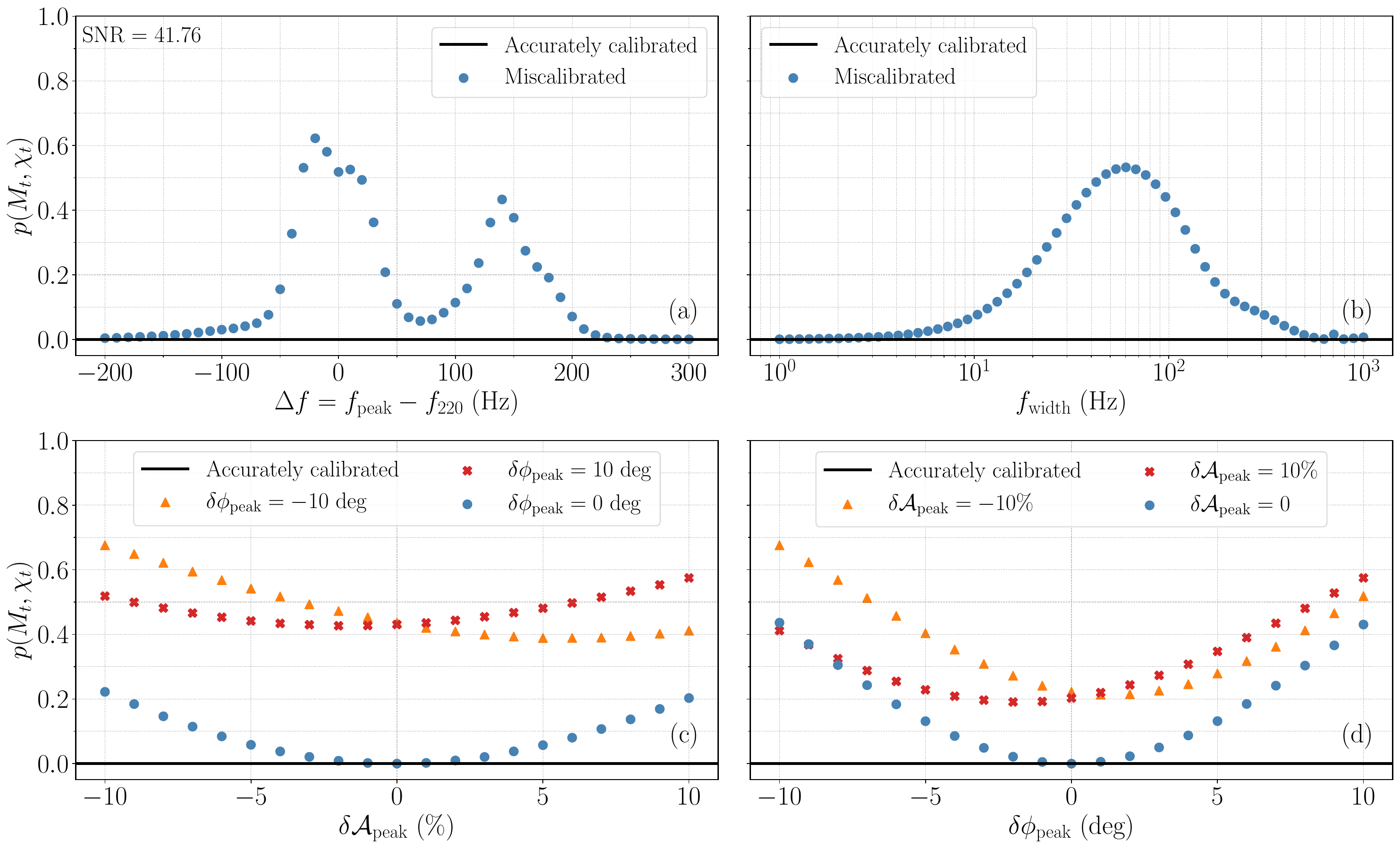}%
    \caption{\label{fig:nonoise330}(Similar to Fig.~\ref{fig:pval330}; see details in Fig.~\ref{fig:pval330} caption)Joint posterior quantile values, $p(M_t, \chi_t)$, for a manually constructed 220 and 330 signal injected in aLIGO Hanford without additive noise as a function of the calibration error properties. 
    }
\end{figure*}

We conduct a similar validation using a manually-constructed signal that includes the 220 and 330 QNMs, shown in Fig.~\ref{fig:nonoise330}. Generally, the differences between Figs.~\ref{fig:nonoise330} and \ref{fig:pval330} are as expected and similar to the differences discussed above. Notably, the second peak in panel (a), where we vary $\Delta f$ around the 330 QNM, is larger in Fig.~\ref{fig:nonoise330} (the controlled study). This occurs because the relative strength of the 330 QNM in the NR waveform is smaller.

\bibliography{final}

\end{document}